\documentclass[twocolumn]{article}
\pdfoutput=1
\usepackage{cite}
\usepackage{amsmath}
\usepackage{color,colortbl}
\usepackage[pdftex]{graphicx}

\begin{document}

\title{Resilient Work Stealing}

\author{
Costanza, Pascal\\
ExaScience Lab, imec Belgium\\
\texttt{pascal.costanza@imec.be}
\and
Herzeel, Charlotte\\
ExaScience Lab, imec Belgium\\
\texttt{charlotte.herzeel@imec.be}
\and
De Meuter, Wolfgang\\
Software Languages Lab, Vrije Universiteit Brussel, Belgium\\
\texttt{wdmeuter@vub.ac.be}
\and
Wuyts, Roel\\
ExaScience Lab, imec Belgium\\
\texttt{roel.wuyts@imec.be}
}

\date{March 12, 2013}
\maketitle

\begin{abstract}
Future generations of processors will exhibit an increase of faults over their lifetime, and it becomes increasingly expensive to solve the resulting reliability issues purely at the hardware level. We propose to model computations in terms of restartable task graphs in order to improve reliability at the software level. 
As a proof of concept, we present \emph{Cobra}, a novel design for a shared-memory work-stealing scheduler that realizes this notion of restartable task graphs, and enables computations to survive hardware failures due to soft errors. A comparison with the work-stealing scheduler of Threading Building Blocks on the PARSEC benchmark suite shows that Cobra incurs no performance overhead in the absence of failures, and low performance overheads in the presence of single and multiple failures.
\end{abstract}

\section{Introduction}
\label{sec:introduction}

Due to extreme reduction in scale, future generations of processors will exhibit an increase of software-visible failures, and handling such failures at the hardware level becomes increasingly expensive in terms of hardware complexity and energy consumption~\cite{borkar05,dongarra11,elnozahy08,rivers09}. Therefore, it is necessary to investigate how existing hardware solutions can be complemented at the software level~\cite{debardeleben09,elnozahy08,ferreira11,shye09}.

Software solutions for such reliability issues already exist primarily in the field of high-performance computing, where the state of the art is based on checkpointing strategies. However, checkpointing alone is predicted to become infeasible in the future because of the large overhead and high checkpoint frequency necessitated by ever decreasing mean times between failures~\cite{bosilca12,cappello09,dongarra11,elnozahy08}.

As a novel approach, we propose to model computations in terms of restartable task graphs to improve reliability at the software level, above the hardware, but below a more costly checkpointing infrastructure. A failure that occurs during the execution of a task graph can be isolated by identifying a corresponding subgraph whose re-execution fixes the cause of the failure and allows the remaining task graph to be completed successfully. In this way, the failure can be isolated and its impact on the running software minimized. Only if a failure-correcting subgraph cannot be identified, does the failure need to be reported to the next level of the software resilience stack, for example as a trigger for a checkpoint rollback recovery.

As a proof of concept that realizes this notion of restartable task graphs we present \emph{Cobra}, a novel design for a shared-memory work-stealing scheduler. 
Unlike traditional work-stealing approaches, Cobra maintains an explicit representation of the tree-shaped graph generated by a fork/join computation. In the presence of \emph{soft errors}, it can selectively re-execute tasks to identify a failure-correcting subtree that implicitly fixes the cause of the failure, provided computations are expressed as \emph{idempotent} algorithms. A comparison with Threading Building Blocks on the PARSEC benchmark suite shows that a C++11 implementation of Cobra incurs no performance overhead in the absence of failures, and low performance overheads in the presence of single and multiple failures.

\section{Restartable task graphs for resilience}
\label{sec:restartable-task-graphs}

\subsection{Faults and failures}
\label{sec:faults-and-failures}

\emph{Faults} are anomalies at the hardware level that may or may not give rise to \emph{failures}, such as \emph{silent errors}, \emph{crashes} or \emph{hangs}. For example, alpha particles and cosmic rays may cause spontaneous bit flips in memory, and may give rise to \emph{soft errors}, i.e., wrong memory contents, which may cause computations to return incorrect results, make pointers refer to illegal memory addresses, or invalidate termination conditions in loops~\cite{mukherjee08}. \emph{Error-correcting code memory} (ECC) can prevent soft errors, but only up to a certain number of simultaneous bit flips in a memory region. Since in future hardware, the number of bit flips is expected to rise significantly due to extreme reduction in scale, the rate of software-visible failures will increase, and thus software solutions to improve resilience become necessary~\cite{borkar05,dixit11}.

\subsection{Requirements}
\label{sec:restartable-tasks-requirements}

The goal of our approach is to isolate failures and to minimize their impact on the running software as much as possible. Our key insight is that we need a runtime infrastructure with a light-weight mechanism that restarts only a relatively small number of isolated, fine-grained software tasks as a reaction to a failure. This contrasts with checkpoint/restart techniques where with \textit{uncoordinated} strategies at least one or more, and with \textit{coordinated} strategies even all involved compute nodes need to be restarted from a checkpoint previously written to secondary storage, which implies high costs in terms of writing checkpoints, downtime, coordination, and recovery.

We can envision several options for runtime infrastructures that allow for fine-grained tasks to be restarted. We propose the following design space to differentiate between the possible solutions.

\subsubsection{Explicit tasks and dependencies} In order to know what tasks to restart in the presence of failures, these tasks have to be explicitly represented, or it must be possible to reliably reconstruct them. Moreover, if a task that needs restarting impacts other, dependent tasks, such dependencies need to be explicitly represented, or reliably reconstructable, as well.

\subsubsection{Fault-detection mechanism} Task restarts need to be triggered when a hardware failure occurs, and there must be a minimum likelihood that such restarts fix the cause of the failure, so that the computation has a chance to successfully complete. Therefore, a mechanism is needed that enables software to both recognize and fix failures. This mechanism depends on the fault model one wishes to tackle. In this paper, we focus on soft errors, especially multi-bit flips in memory. Other hardware faults that can be fixed through task restarts may include communication faults, logic faults, and so on.

\subsubsection{Task restart mechanism} A mechanism is needed that manages the task restarts. This includes identifying the task and/or subgraph of the task graph that needs to be restarted in order to fix the cause of a particular hardware failure. For example with soft errors, a failure may manifest itself in the form of a segmentation fault when dereferencing an invalid pointer, but the task that can fix the cause of the failure by assigning a valid address to the pointer variable may only occur higher up in the task graph. Tasks that depend on the identified task to be restarted may be in need of restarting as well as a consequence of that restart. It must also be ensured that task restarts do not alter the result of the overall computation. This is a non-trivial requirement, especially when task graphs can modify (i.e., read and write, instead of only either read or write) shared memory.

\subsection{The Cobra Approach}
\label{sec:the-cobra-approach}

The requirements stated above for restartable task graph runtimes leave a lot of room for concrete realizations. In principle, it would be desirable to design such an infrastructure for task graphs with \emph{arbitrary} dependencies. However, it is known that scheduling arbitrary task graphs is NP-complete, even if the task graphs are static and compute times are known in advance~\cite{ullman75}. Although efficient heuristics and approximations exist for static task graphs with known compute times~\cite{kwok99}, we expect that restarting tasks and/or subgraphs due to hardware failures at runtime will have a non-trivial impact on such schedulers for arbitrary task graphs.

On the other hand, \emph{fully strict} computations, which can be expressed with fork/join primitives, yield tree-shaped graphs for which \emph{work stealing} is known to create efficient schedules on the fly, with provably optimal time and space bounds~\cite{blumofe99}. In practice, work stealing has been shown to be effective for example for spawn/sync in Cilk~\cite{frigo98}, spawn/wait in Threading Building Blocks~(TBB)~\cite{reinders07}, untied tasks in OpenMP~\cite{duran08}, Java fork/join~\cite{lea00}, and async/finish in Habanero and X10~\cite{charles05}. We therefore opted to initially focus on the combination of fork/join computations with work stealing as well.

The result is \emph{Cobra}, a novel resilience-aware work-stealing scheduler for fully strict tree-recursive fork/join computations. The remainder of this section shows how Cobra realizes the above requirements, while Section~\ref{sec:design} describes Cobra in more detail.

\subsubsection{Explicit tasks and dependencies} Existing work-stealing schedulers optimize away the tree structure of fork/join computations as much as possible, by removing tasks from work-stealing queues as early as possible, and using the native execution stack as far as possible. As a consequence, tasks that need to be restarted because of failures are usually already gone, and reconstructing them is difficult. Cobra represents every task by a node in an explicit representation of the fork/join tree instead, which maintains references to its children and to its join continuation. This tree data structure differs substantially from implicit task representations in already existing work-stealing schedulers.

\subsubsection{Fault-detection mechanism} Recent processors by Intel and AMD have integrated a \emph{machine check architecture}, which reports corrected, fatal, and software-recoverable faults to the operating system~\cite{amd12,intel11}. In the case of soft errors, the machine check architecture can mark memory with bit flips that ECC memory cannot recover from as \emph{poisoned}. If the software tries to consume poisoned memory, the hardware throws a \emph{machine check exception} instead, and by doing so, gives the operating system a chance to recover. For example, Linux since kernel version 2.6.x deals with some of these hardware-level exceptions transparently, for example by invalidating caches that contain poisoned data. It also reports some of these exceptions as signals to processes when they attempt to consume poisoned memory, and thus gives applications a chance to recover in turn~\cite{kleen09,kleen10}.

Cobra uses such data poisoning as its fault-detection mechanism. The tree structure of a fork/join computation allows the scheduler to detect a task that fails due to a read access to poisoned memory, and to trigger its re-execution. If that task performs a write access to the poisoned memory location, and thus ``unpoisons'' it before another read access to the same memory location occurs, the remainder of the computation can succeed. If that task continues to fail, the scheduler can select nodes higher up in the tree for re-execution, until the necessary unpoisoning write access is performed by a subtree that allows the overall computation to succeed.\footnote{The current design of the machine check architecture for x86 processors does not support unpoisoning of memory locations by itself, because it typically marks larger regions, of the size of a cache line or memory page, as poisoned, and does not support partially poisoned regions. However, a separate layer on top of the machine check architecture, which remaps poisoned memory and intercepts memory accesses to such regions by way of memory page protection, can provide the necessary unpoisoning of memory in software, without affecting accesses to other, unpoisoned regions.}

A fork/join computation may not unpoison a poisoned memory location, that is, it may not write to such a memory location anymore after it gets corrupted. If the computation also does not read from such a poisoned memory location anymore either, the soft error \emph{escapes} from the fork/join computation. In that case, Cobra cannot recover, and later accesses to escaped poisoned memory locations need to rely on a more costly resilience mechanism higher up in the software stack, such as checkpointing. However, Cobra may still significantly reduce the number of soft errors that escape from fork/join computations, which can help to reduce the frequency, and thus the cost of checkpointing. 

\subsubsection{Task restart mechanism} 
In work-stealing schedulers, task execution is distributed over several worker threads that operate on the fork/join computation in parallel by executing the same scheduling loop. Cobra integrates the monitoring and rescheduling of failed nodes as part of the scheduling loop. We discuss the details in Section \ref{sec:design}. 

To ensure that restarts do not alter the result of the overall computation expressed by a fork/join tree, Cobra requires tasks to be expressed as idempotent algorithms, where identifiable steps may be executed more than once without producing different results than when executed exactly once. The primary reason for an algorithm not to be idempotent is when it uses assignment statements that both read from and write to the same memory location. For example, \texttt{x = x + 1} is not idempotent, because \texttt{n} executions will increment \texttt{x} by \texttt{n}, not by \texttt{1} as intended. Purely functional programs are by definition idempotent, but a wide class of imperative algorithms, such as the benchmarks we use to validate Cobra in Section~\ref{sec:benchmarks}, can also be expressed in an idempotent style. This is also in line with previous observations~\cite{kruijf12,michael09}.

\section{Design of Cobra}
\label{sec:design}

This section discusses the design of Cobra: how the fork/join tree is represented; how the scheduling loop executes user code and enables it to dynamically build the fork/join tree; and how the scheduling loop monitors failed tasks and restarts subtrees to fix such failures.

\subsection{Representation}
\label{sec:design-fork-join}

\begin{figure}
\begin{footnotesize}
\begin{verbatim}
struct node_type {
  parent;
  children;
  task;
  continuation;
  state;
}

void fork(node, task) {
  child              = new node_type;
  child.parent       = node;
  child.task         = task;
  child.children     = empty;
  child.continuation = empty;
  child.state        = free;
  node.children.add(child);
}

void join(node, continuation) {
  node.continuation = continuation;
}
\end{verbatim}
\end{footnotesize}
\caption{Representation of \texttt{nodes} in Cobra.}
\label{code:continuator}
\end{figure}

In Cobra, the fork/join tree is created from nodes. Fig.~\ref{code:continuator} shows an abstract definition of such nodes. Each node has the following attributes:
\begin{itemize}
\item a reference to its \emph{parent} in the tree (the root node has an empty \emph{parent} reference);
\item a \emph{children} container with references to one or more child nodes (leaves have empty \emph{children} containers);
\item \emph{task}/\emph{continuation} references provided by user code;\footnote{Cobra uses lambda expressions to represent tasks and continuations, but task/continuation objects like in Java fork/join or TBB could also be used.}
\item a \emph{state}, which can be one of \emph{free}, \emph{busy}, and \emph{done}, and in the presence of failures, can additionally be \emph{inactive}.
\end{itemize}

The fork/join tree gets constructed during execution of an algorithm. The operation \emph{fork} (Fig.~\ref{code:continuator}) creates a new child node; assigns the \emph{parent} and \emph{task} references as provided by arguments; initializes \emph{children} and \emph{continuation} to \emph{empty}, and \emph{state} to \emph{free} as default values; and adds the new child to the \emph{children} container of the \emph{parent} node. The operation \emph{join} just assigns the \emph{continuation} attribute.

 Fig.~\ref{code:parallel_quicksort} shows pseudo-code for a parallel quicksort in terms of \emph{fork}/\emph{join} operations as an example: The function \emph{qsort} operates on an array on the region defined by \emph{start} and \emph{end}. If the region is not empty, it partitions the array and determines a new pivot element. It then forks invocations of \emph{qsort} for the left and the right partitions, which can be executed in parallel. It finally specifies an invocation of \emph{merge} as the join continuation.\footnote{Definitions for \emph{partition} and \emph{merge} are not shown here. We leave it open how the \emph{current} node is determined on which \emph{fork} and \emph{join} are invoked. For example, the \emph{current} node can be passed as an argument to a forked lambda expression, or it can be bound to a \emph{thread-local} variable.}

\begin{figure}
\begin{footnotesize}
\begin{verbatim}
void qsort(array, start, end) {
  if (start < end) {
    pivot = partition(array, start, end));
    fork(node, qsort(array, start, pivot));
    fork(node, qsort(array, pivot+1, end));
    join(node, merge(array, start, pivot, end));
  }
}
\end{verbatim}
\end{footnotesize}
\caption{Parallel \texttt{quicksort} in terms of fork/join.}
\label{code:parallel_quicksort}
\end{figure}

\begin{figure}
\begin{center}
\includegraphics[scale=0.46]{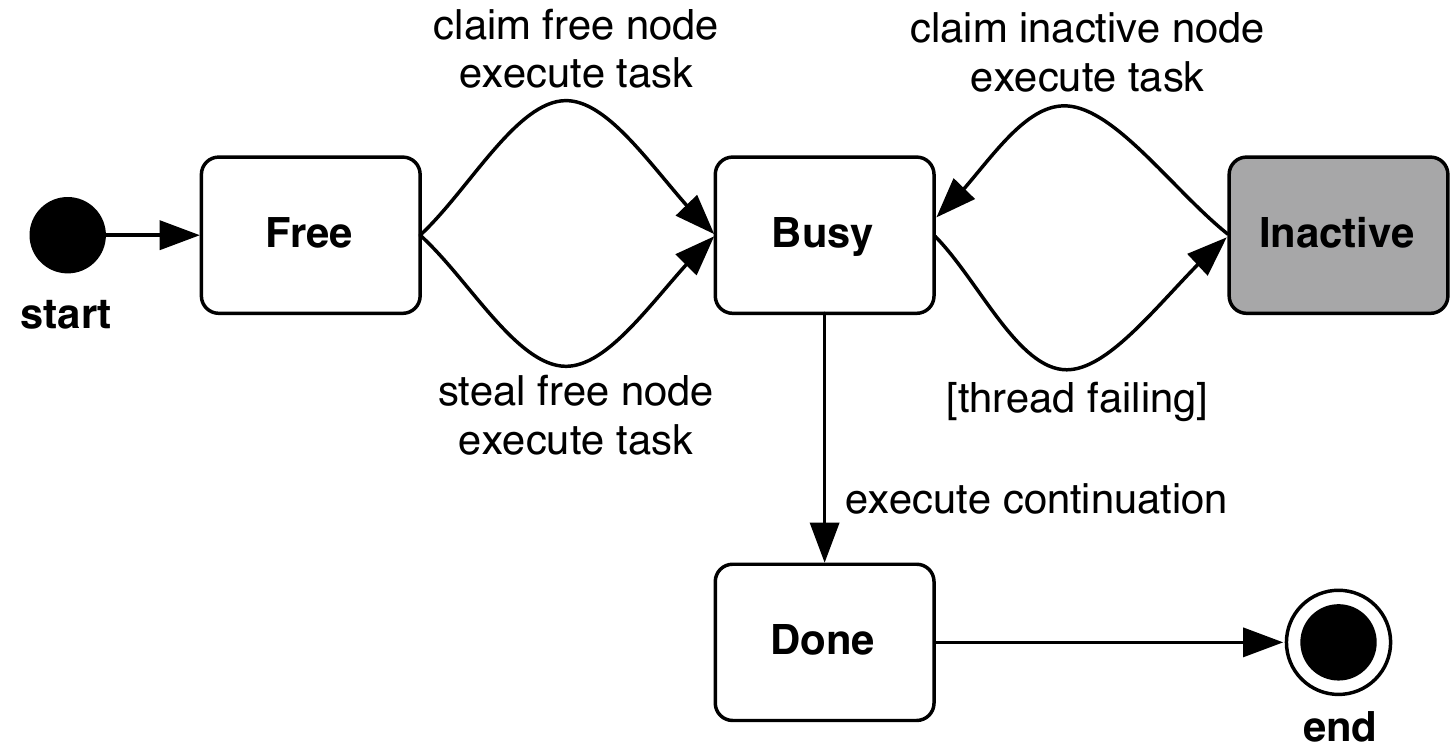}
\end{center}
\caption{The life cycle of a Cobra node.}
\label{fig:node-states}
\end{figure}

Note that \emph{fork} and \emph{join} do not immediately execute user code provided in \emph{task} and \emph{continuation}. This is the responsibility of the scheduling loop, detailed in Section~\ref{sec:scheduling-loop}.

The scheduling loop is also responsible for performing the \emph{state} transitions in a node, depicted in Fig.~\ref{fig:node-states}: Each node starts in \emph{state free}. When a worker thread claims or steals a node, it executes its \emph{task}, creating its \emph{children} and \emph{continuation}, if any, and then transitions its \emph{state} to \emph{busy}. In failure-free execution, the \emph{continuation} of a node is executed when all its \emph{children} are \emph{done}, and its \emph{state} then transitions to \emph{done} as well. In the presence of failures, the \emph{state} of a node may transition from \emph{busy} to \emph{inactive}. It can then be reclaimed, which eventually transitions its \emph{state} back to \emph{busy}.


\subsection{Failure-free execution}
\label{sec:scheduling-loop}

In Cobra, like in other work-stealing schedulers, the scheduling and execution of tasks and continuations is distributed over several worker threads. Each thread owns a working list of nodes that it is responsible for, and executes the same scheduling loop. Unlike in other work-stealing schedulers, the nodes that are stored in Cobra's working lists are already claimed by the corresponding worker thread and cannot be stolen by other threads. Only \emph{children} of such nodes can be claimed or stolen in Cobra.

\begin{figure}[!t]
\includegraphics[scale=0.4]{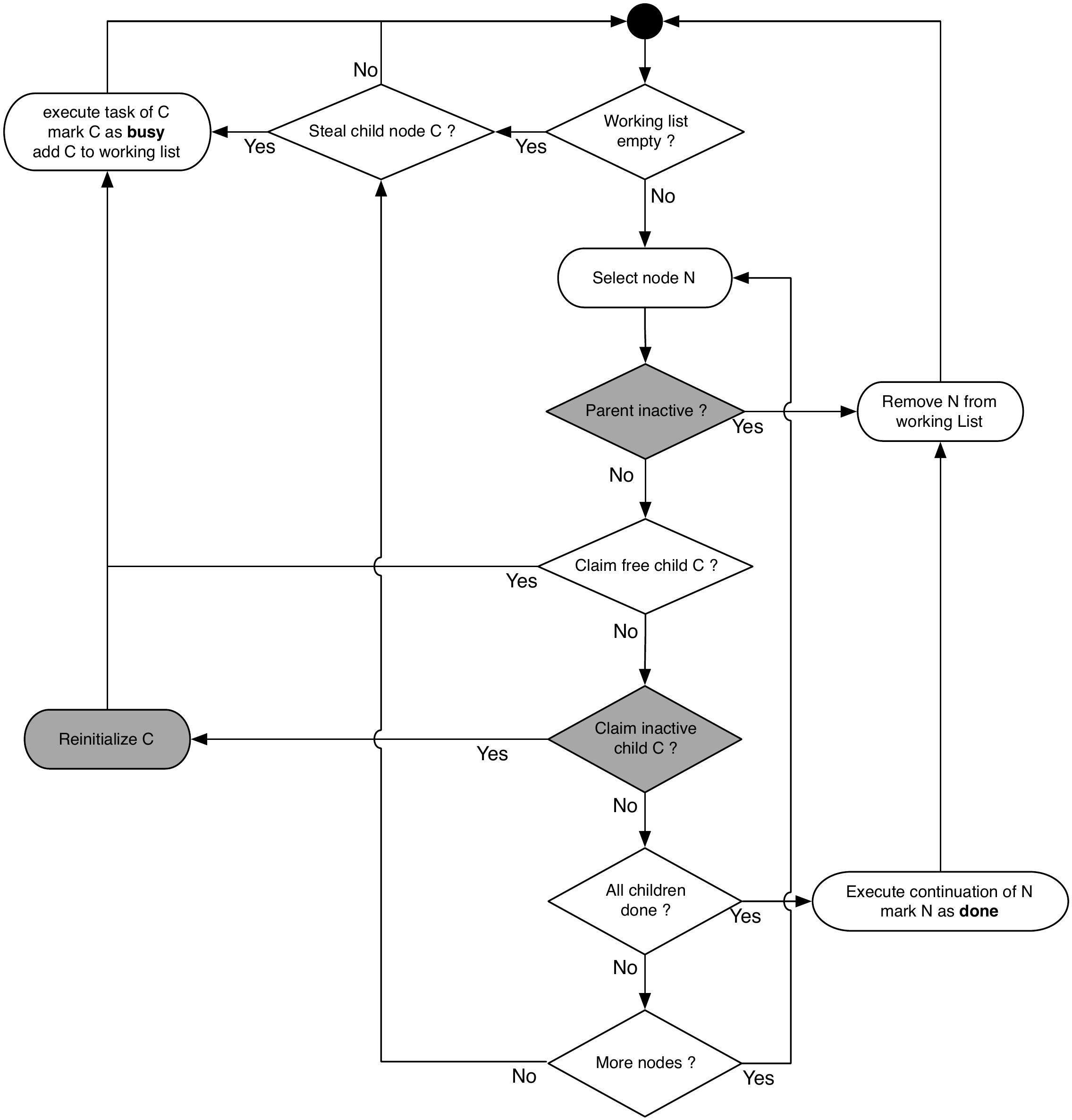}
\caption{The scheduling loop of Cobra. Grey boxes deal with nodes that are inactive due to failures.}
\label{fig:scheduling-loop-diagram}
\end{figure}

The flow of the scheduling loop in each worker thread is depicted in Fig.~\ref{fig:scheduling-loop-diagram}, with the filled circle being the starting point. Assume that a fork/join computation starts with one working list containing the root node, and all other working lists being empty. Further assume that the root node's \emph{task} is already executed, and thus the \emph{children} references and \emph{continuation} of the root node are already defined.

If the working list of the current worker thread is empty, the thread attempts to steal a node from another worker thread (see below). If the working list is not empty (for example, when it contains the root node), the worker thread selects a node \emph{N} from its working list. 

If the \emph{parent} of \emph{N} is \emph{inactive}, the node is removed from the working list and the scheduling loop is restarted. In failure-free execution, nodes are never \emph{inactive}, so we skip this case until Section~\ref{sec:design-monitoring}.

If a child \emph{C} of \emph{N} is in state \emph{free}, the worker thread can claim it, execute its \emph{task}, set its \emph{state} to \emph{busy}, and enter it into its working list. The scheduling loop is then restarted. If no child of \emph{N} is \emph{free}, the worker thread tries to claim an \emph{inactive} child instead, which we discuss in the next subsection. 

If no child of \emph{N} can be claimed, then either all \emph{children} are \emph{done}, or some \emph{children} are \emph{busy}. If all \emph{children} are \emph{done} (including when there are no \emph{children}), the worker thread executes the \emph{continuation} of \emph{N}, removes \emph{N} from the working list, sets the \emph{state} of \emph{N} to \emph{done}, and restarts the scheduling loop. If not all \emph{children} are \emph{done}, which means that some \emph{children} are \emph{busy}, and there are more other nodes in the working list, another node \emph{N} is selected for inspection. If after inspecting all nodes in the working list, the scheduling loop has not been restarted, then all nodes in the working list have \emph{busy children}, and the worker thread attempts to steal a node from another worker thread.

When a worker thread attempts to steal a node, it chooses another victim thread at random, and searches for a \emph{free} child of any node in the victim's working list. A steal attempt may fail, when a victim's working list contains no nodes or no \emph{children}, or when no child of any node in the victim's working list is \emph{free}. In case the thief's working list is empty, the thief will make more steal attempts, potentially from other victims, until it successfully steals a node. If the thief's working list is not empty, it continues by inspecting the nodes in its own working list, to ensure their progress. A worker thread attempts to steal nodes even if its own working list is not empty, to improve overall progress and avoid idling as much as possible when all \emph{children} of all nodes in its own working list are \emph{busy}.

Observe that only a \emph{busy} node can enter a working list, after its \emph{task} has been executed, and therefore its \emph{children}, if any, and its \emph{continuation}, if any, have been created. Furthermore, a node is only removed from a working list after its \emph{children} are all \emph{done}, its \emph{continuation} has been executed, and its \emph{state} has been set to \emph{done} as well.

\subsection{Distributed node monitoring}
\label{sec:design-monitoring}

The previous subsection focuses on the failure-free execution of the scheduling loop in Fig.~\ref{fig:scheduling-loop-diagram}. As outlined in Section~\ref{sec:the-cobra-approach}, Cobra's fault model is based on soft errors, and more specifically on data poisoning: When either user code or a worker thread itself attempts to consume poisoned memory, the thread executing that attempt receives a machine check exception. In reaction, it sets the \emph{state} of all nodes in its working list to \emph{inactive}, reinitializes its working list and other data structures relevant for the scheduler, and restarts its scheduling loop with an empty working list. As a consequence, nodes that become \emph{inactive} due to such a failure are not referenced in any working list anymore and need to be reclaimed. We decided against keeping track of the particular node whose execution caused the failure in order not to penalize computations that do not encounter any soft errors, which are still rare events.

The flow diagram in Fig.~\ref{fig:scheduling-loop-diagram} depicts three boxes that deal with \emph{inactive} Cobra nodes, highlighted in grey:
\begin{itemize}
\item After a node \emph{N} is selected for inspection when a working list is not empty, it is checked whether the state of the node's \emph{parent} reference has become \emph{inactive} as a reaction to a machine check exception. In that case, \emph{N} is no longer relevant because it will eventually be recreated as part of the re-execution of the top-most \emph{inactive parent} of \emph{N}.
\item If a worker thread cannot claim a \emph{free} child of \emph{N} in the next step, it then checks whether a child \emph{C} of \emph{N} is in state \emph{inactive}. If so, it claims it, reinitializes its \emph{children} and \emph{continuation} to \emph{empty}, re-executes its task, sets its \emph{state} to \emph{busy}, and enters it into its working list. Recall that executing a task allows user code to invoke \emph{fork} and \emph{join} in order to redefine \emph{children} and \emph{continuation}, which implictly reconstructs the failed subtree. The scheduling loop is then restarted.
\end{itemize}

We do not allow an \emph{inactive} node to be stolen. This ensures that it can only be restarted by the thread working on its \emph{parent} node. If this does not unpoison the problematic memory location, the restarted node is likely to fail again. This triggers a failure in that worker thread, and thus sets the \emph{state} of the \emph{parent} node to \emph{inactive} as well, which in turn can only be restarted by the thread working on the \emph{parent} node's \emph{parent}, and so on. This ensures that a failure percolates up the fork/join tree until it is either fixed, if necessary by restarting the complete tree from its root node; or else, it gets reported as a failure of the overall computation, to be handled by a more costly resilience mechanism higher up in the software stack. Note that an \emph{inactive} node is not claimed again in case its \emph{parent} node is also \emph{inactive}. In that case, there is, however, a top-most \emph{inactive} node in the transitive closure of the \emph{parent} relation, which is eventually claimed by the thread working on its \emph{active} parent.

Cobra is effectively a trampoline-style scheduler: Every invocation of user code through \emph{fork} and \emph{join} returns to the scheduling loop, which can in turn execute available nodes, execute enabled join continuations, check for liveness of parents and children of nodes stored in its working list, drop children of inactive parents, and reschedule inactive child nodes.

Cobra survives failures that occur in user code because nodes are only modified when worker threads are responsible for them after successful claim or steal attempts. The node's \emph{state} is eventually changed to \emph{done}, but only when it is completely finished, after executing its continuation. If the thread responsible for the node fails at any stage before, the \emph{state} is not set to \emph{done}, and the worker thread responsible for the node's \emph{parent} can take over.


\subsection{Implementation details}
\label{sec:implementation}

In Cobra, multiple worker threads may attempt to access the same Cobra node in parallel. This requires synchronization to avoid data races. To avoid unnecessary locking overheads, our implementation is based on a lock-free protocol using atomic compare-and-swap operations. This is possible because of the \emph{partial monotonicity} of state transitions in Cobra nodes: A node can only ever change from \emph{free} to \emph{busy}, from \emph{busy} to \emph{inactive} and back in the presence of failures, and from \emph{busy} to \emph{done} (Fig.~\ref{fig:node-states}).

The simplest case is when a worker thread tries to claim a \emph{free} node locally which, when successful, changes the \emph{state} of the node to \emph{busy}. This is expressed in the Cobra implementation by way of an atomic compare-and-swap operation (CAS), as shown in the pseudo-code in Fig.~\ref{code:try_free_node}. The operation \texttt{try\_free\_node} checks whether a node is \emph{free} and in case it is, performs the transition to \emph{state} \emph{busy} in one atomic step. This use of an atomic compare-and-swap operation prevents more than one thread from successfully claiming the same \emph{free} node at the same time. In case the comparison fails, another thread has already claimed the node, or the node may even already be \emph{done}. In case \texttt{try\_free\_node} successfully claims a node, it executes the node's \emph{task}, filling the node's \emph{children} and \emph{continuation} attributes, if any, and reports to the scheduling loop that the claim is indeed successful.

\begin{figure}
\begin{footnotesize}
\begin{verbatim}
boolean try_free_node (node) {
  if (CAS(node.state, free => busy)) {
    execute node.task;
    return true;
  }
  return false;
}
\end{verbatim}
\end{footnotesize}
\caption{Locally claiming a \emph{free} Cobra node.}
\label{code:try_free_node}
\end{figure}

Another way to claim a node locally is when a worker thread tries to claim it as an \emph{inactive} node. The corresponding operation \texttt{try\_inactive\_node} is shown in Fig.~\ref{code:try_inactive_node}.  It first checks whether the given node is \emph{inactive} and if this is the case, also performs a transition to \emph{state} \emph{busy}. Here, the check and \emph{state} transition is not combined into an atomic operation, because the only thread trying to claim an \emph{inactive} node is the worker thread that is responsible for that node's \emph{parent}, to ensure percolation of repeated failures. The \emph{state} transition is performed by assignment instead, followed by a reinitialization of the \emph{children} and \emph{continuation} attributes, and execution of the node's \emph{task}.

\begin{figure}
\begin{footnotesize}
\begin{verbatim}
boolean try_inactive_node (node) {
  if (node.state is inactive) { 
    node.version = node.version+1;
    write_barrier;
    node.state = busy;
    node.children = empty;
    node.continuation = empty;
    execute node.task;
    return true;
  }
  return false;
}
\end{verbatim}
\end{footnotesize}
\caption{Locally claiming an \emph{inactive} Cobra node.}
\label{code:try_inactive_node}
\end{figure}

\pagebreak

Recall that we must be able to recognize whether the \emph{parent} of a node is \emph{inactive} or not in the scheduling loop (see Fig.~\ref{fig:scheduling-loop-diagram}). However, the \texttt{try\_inactive\_node} operation may reset the \emph{state} of such a \emph{parent} from \emph{inactive} to \emph{busy}, which may lead a thread operating on the child node to falsely believe that its \emph{parent} is still valid.

To avoid this situation, our Cobra implementation adds a version number as an additional attribute to the node type. The root node is initialized with 0 as its version number, whereas a child node is assigned the version from its \emph{parent} when it is created by a \emph{fork} operation.

When a worker thread successfully claims an \emph{inactive} node, it first increments the version number of that node before it performs any other operations on it (see Fig.~\ref{code:try_inactive_node}). A memory barrier instructs compiler and processor to ensure that these two steps are seen by other threads exactly in that order. This ensures the following invariant: The version of a valid node is always greater than or equal to the version of its \emph{parent}. If a node becomes \emph{inactive}, it can only be claimed in \texttt{try\_inactive\_node}, which transitions its \emph{state} to \emph{busy} and reinitalizes its \emph{children} and \emph{continuation} attributes. Old child nodes that were previously stored in \emph{children} thus either still see an \emph{inactive} \emph{parent} before the memory barrier, or have a version number less than the version of that \emph{parent} after the memory barrier. This allows them to reliably detect their \emph{parent} as invalid in both cases.

\begin{figure}
\begin{footnotesize}
\begin{verbatim}
boolean steal_free_node (node) {
  for each child in node.children {
    if (CAS(child.state, free => busy)) {
      if (node.version <= child.version) {
        execute node.task;
        return true;
      }
    }
  }
  return false;
}
\end{verbatim}
\end{footnotesize}
\caption{Remotely stealing a \emph{free} Cobra node.}
\label{code:steal_free_node}
\end{figure}

The \texttt{try\_free\_node} operation in Fig.~\ref{code:try_free_node} does not need to be adapted for claiming nodes locally due to this introduction of version numbers, because the \emph{parent} node is obviously not \emph{inactive} when its worker thread is performing this operation. For stealing nodes remotely, an additional check is necessary that is shown in Fig.~\ref{code:steal_free_node}. Consider a worker thread attempts to steal a \emph{free} child from a node that has become \emph{inactive}, but is being reactivated, and thus transitions from \emph{inactive} back to \emph{busy} due to the actions of \texttt{try\_inactive\_node} in Fig.~\ref{code:try_inactive_node}. 

There are two cases to consider: Either the steal attempt attempts to claim a child node that it should not, because the reinitialization of the \emph{children} attribute of the \emph{parent} in \texttt{try\_inactive\_node} is not communicated yet; or it does not find any child nodes although it should, because the addition of child nodes to the \emph{children} attribute, which are created when re-executing the node's \emph{task} in \texttt{try\_inactive\_node}, is not communicated yet. In the first case, the version number prevents the found but invalid child from being chosen, and the steal attempt just continues searching for other potential victim nodes; in the second case, the steal attempt completely fails, which just amounts to a longer delay in finding nodes to work on in code that invokes \texttt{steal\_free\_node}. It is known from the literature that steal attempts are rare events in work-stealing schedulers, so the effect of the increase in performance overhead caused by such failed steal attempts is marginal~\cite{frigo98}.

Another important aspect of Cobra is that it can survive not only failures that occur in user code, but also while code of the scheduler itself is being executed. This is ensured by the fact that nodes are only modified when worker threads are responsible for them after successful claim or steal attempts. The \emph{state} of a node is eventually changed to \emph{done}, but only when it is completely finished, after executing its continuation. If the thread responsible for the node fails at any stage before, the node's \emph{state} is not set to \emph{done}, and the worker thread responsible for the node's \emph{parent} can take over. Ensuring that working lists and other internal data structures can always be consistently accessed is similarly straightforward.

\section{Validation and Benchmarks}
\label{sec:benchmarks}

In this section, we discuss the effectiveness and performance of Cobra using the PARSEC~2.1 benchmark suite, a set of multi-threaded programs for shared-memory computers~\cite{bienia08}. We answer two questions:
\begin{enumerate}
\renewcommand{\theenumi}{\Alph{enumi}}
\item How well does Cobra perform?
\item What is the impact of failures?
\end{enumerate}

Our benchmarks consist of the programs in PARSEC~2.1 that are parallelized using the spawn/wait primitives of Threading Building Blocks (TBB): \emph{blackscholes}, a program for calculating the pricing of options using the Black-Scholes partial differential equation; \emph{fluid\-ani\-ma\-te}, a fluid dynamics animation using the smoothed particle hydrodynamics method; \emph{streamcluster}, a kernel for online clustering of an input stream; and \emph{swaptions}, a program for pricing a portfolio of swaptions. 

We modified the programs to run on Cobra by replacing calls to the TBB spawn/wait primitives by equvialent Cobra calls, but left them unchanged in all other respects. For \textit{fluidanimate} and \textit{streamcluster}, we also created idempotent versions of the code. The \textit{blackscholes} and \textit{swaptions} programs are already idempotent in their unchanged versions. The output when running on Cobra was validated by comparing it against the output when running the unchanged programs on TBB. This confirms that running the programs on Cobra produces correct results. All output, and all benchmark numbers presented below, are obtained using the large \emph{native input} data sets of PARSEC~2.1 that resemble real program inputs. 

We ran our experiments on a Ubuntu 10.04 LTS server with four ten-core Intel Xeon E7-4870 processors of each 2.4 GHz and 512 GB of RAM. Cobra is implemented in C++11 and our benchmarks were compiled using gcc-4.6.2.

\subsection{Performance}
\label{sec:overhead-resilience}

\begin{figure*}

\begin{tabular}{ccc}

\includegraphics[scale=0.185]{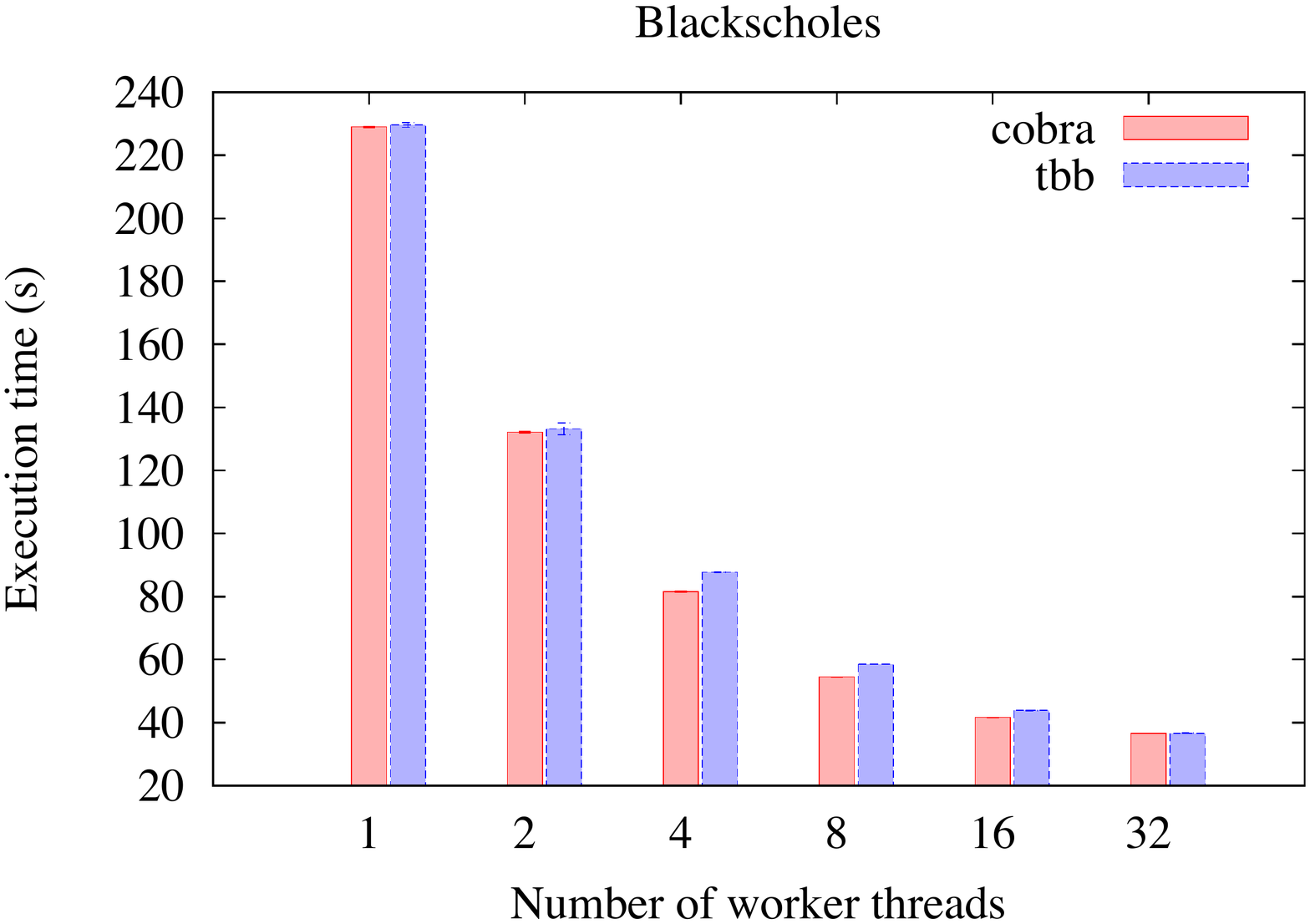}

&

\includegraphics[scale=0.185]{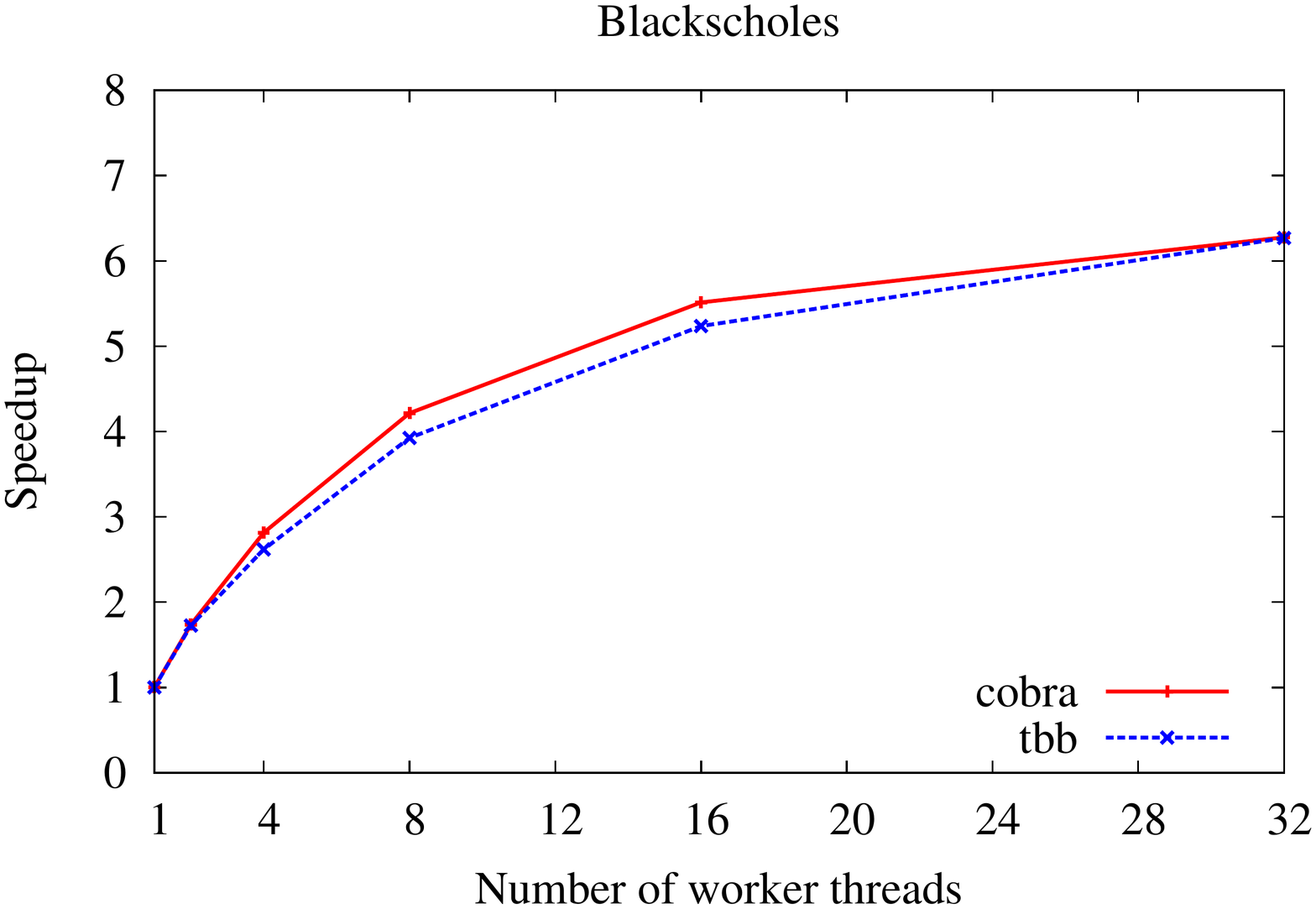}

&

\includegraphics[scale=0.185]{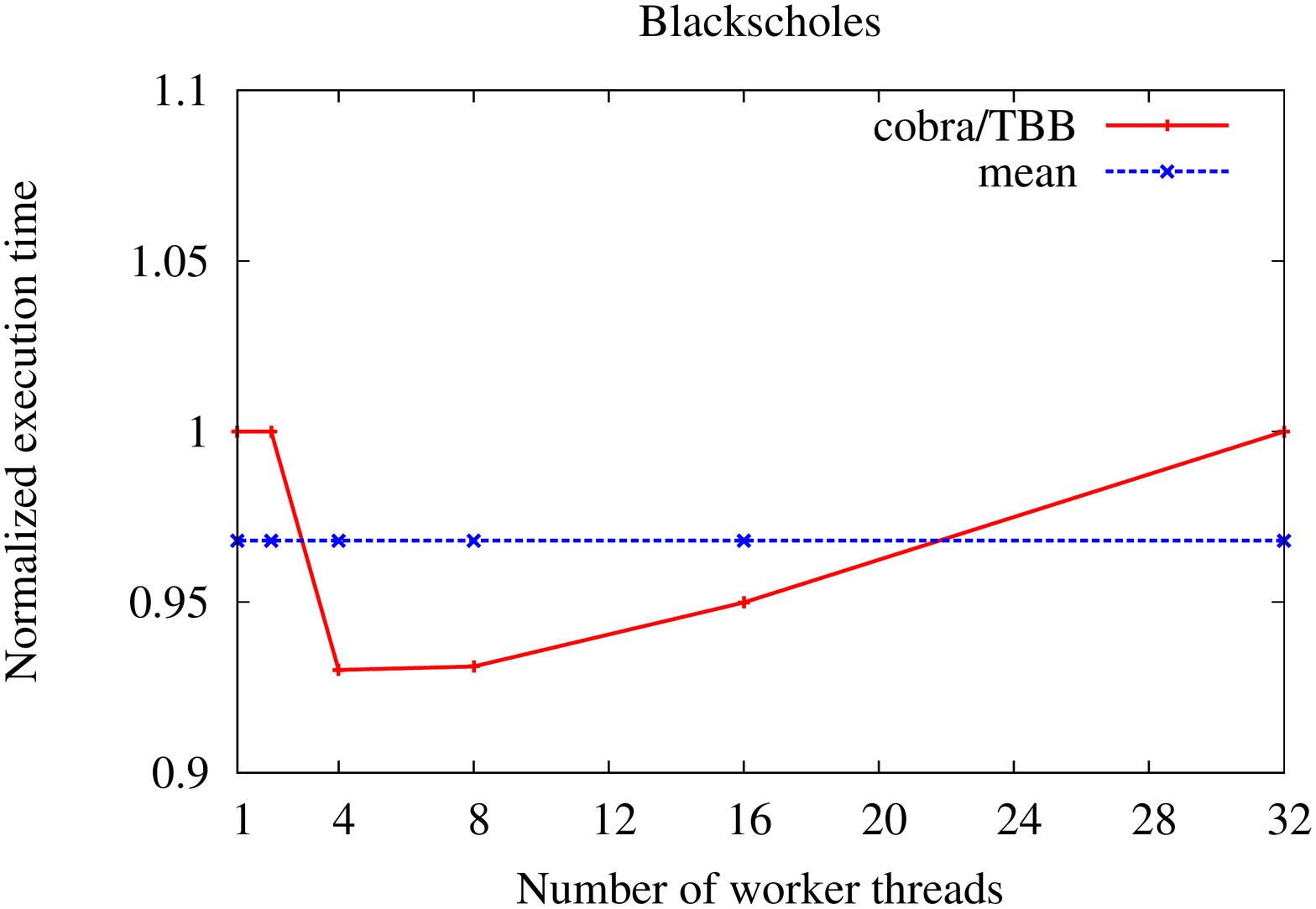}

\\

\includegraphics[scale=0.185]{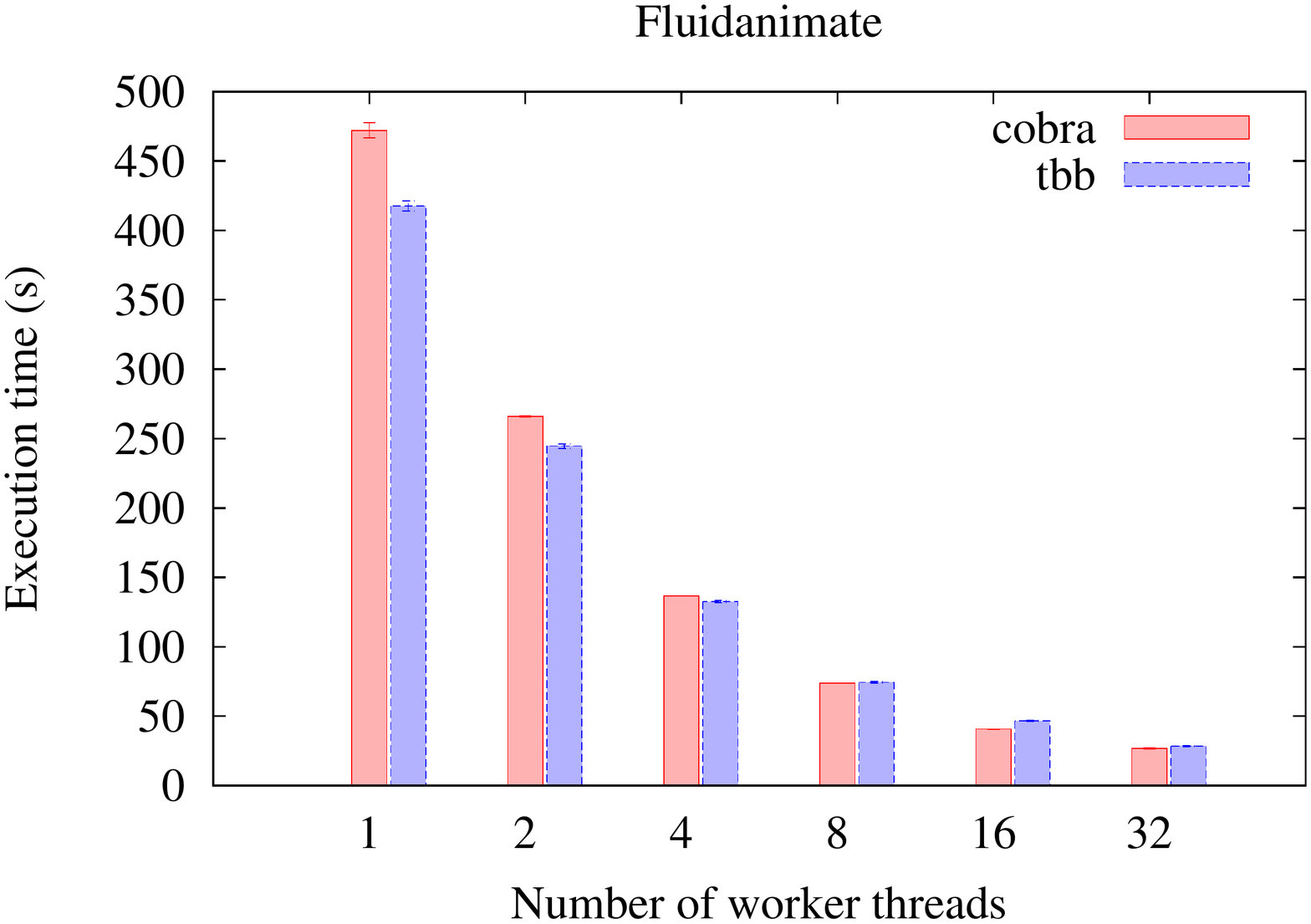}

&

\includegraphics[scale=0.185]{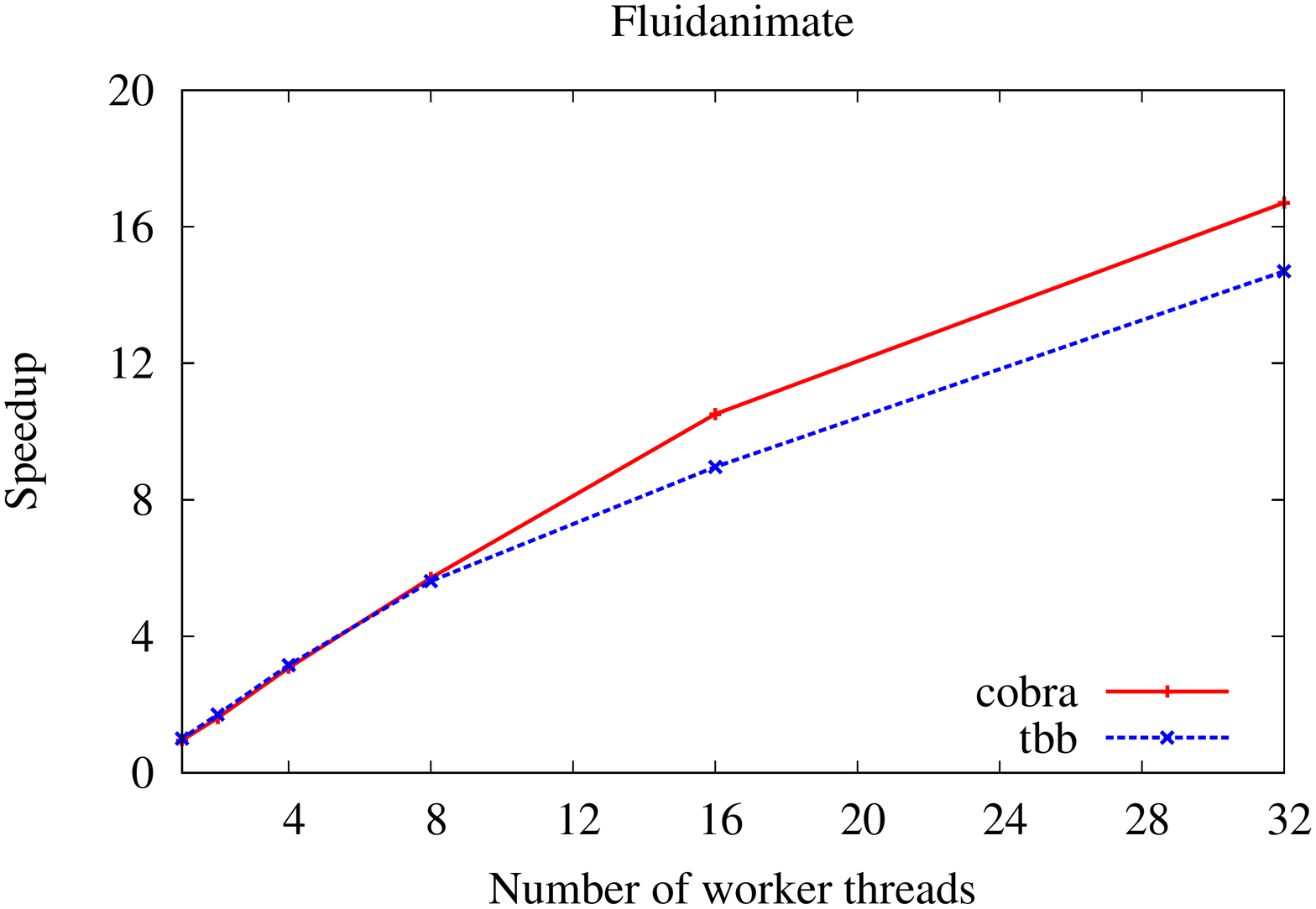}

&

\includegraphics[scale=0.185]{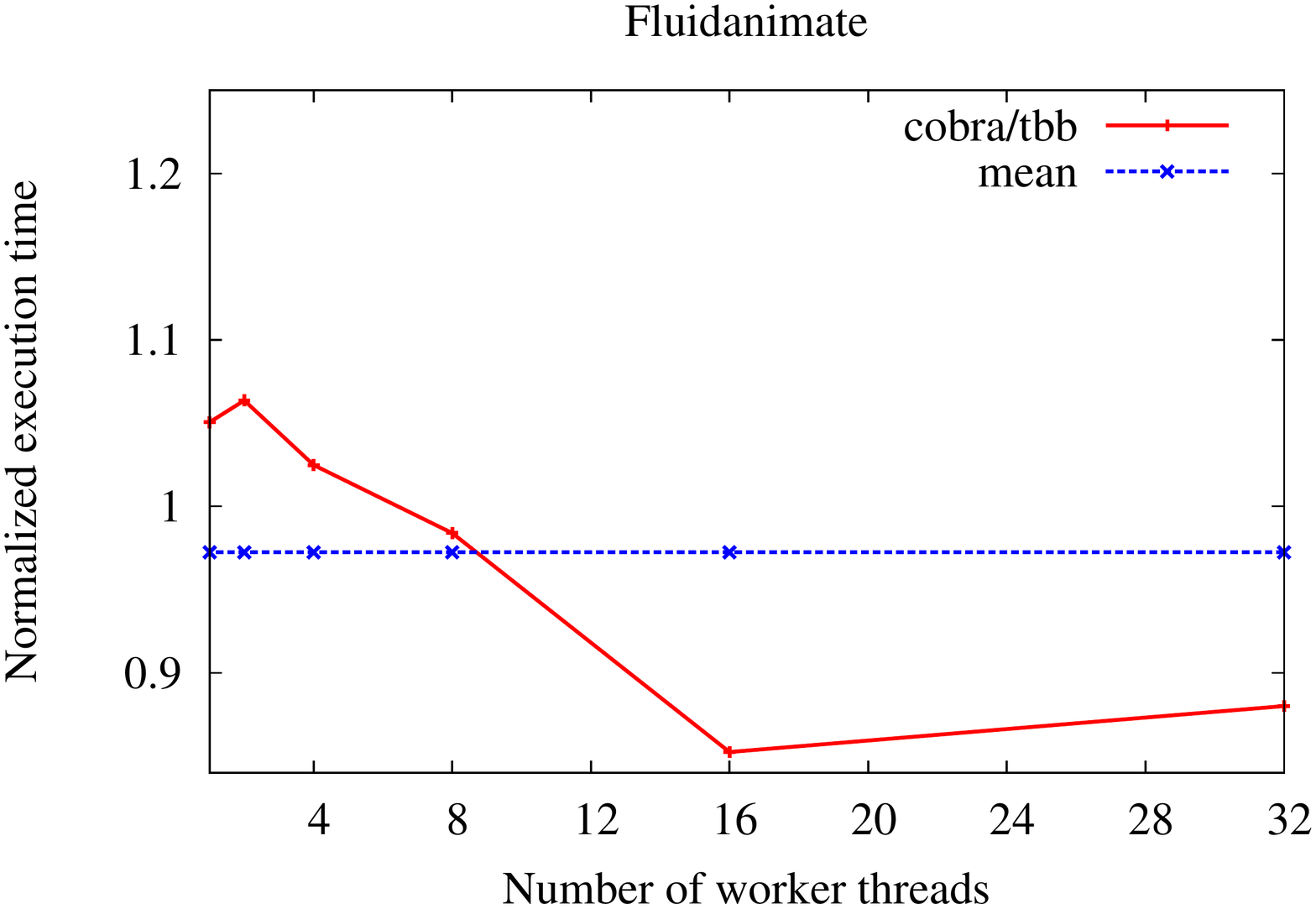}

\\

\includegraphics[scale=0.185]{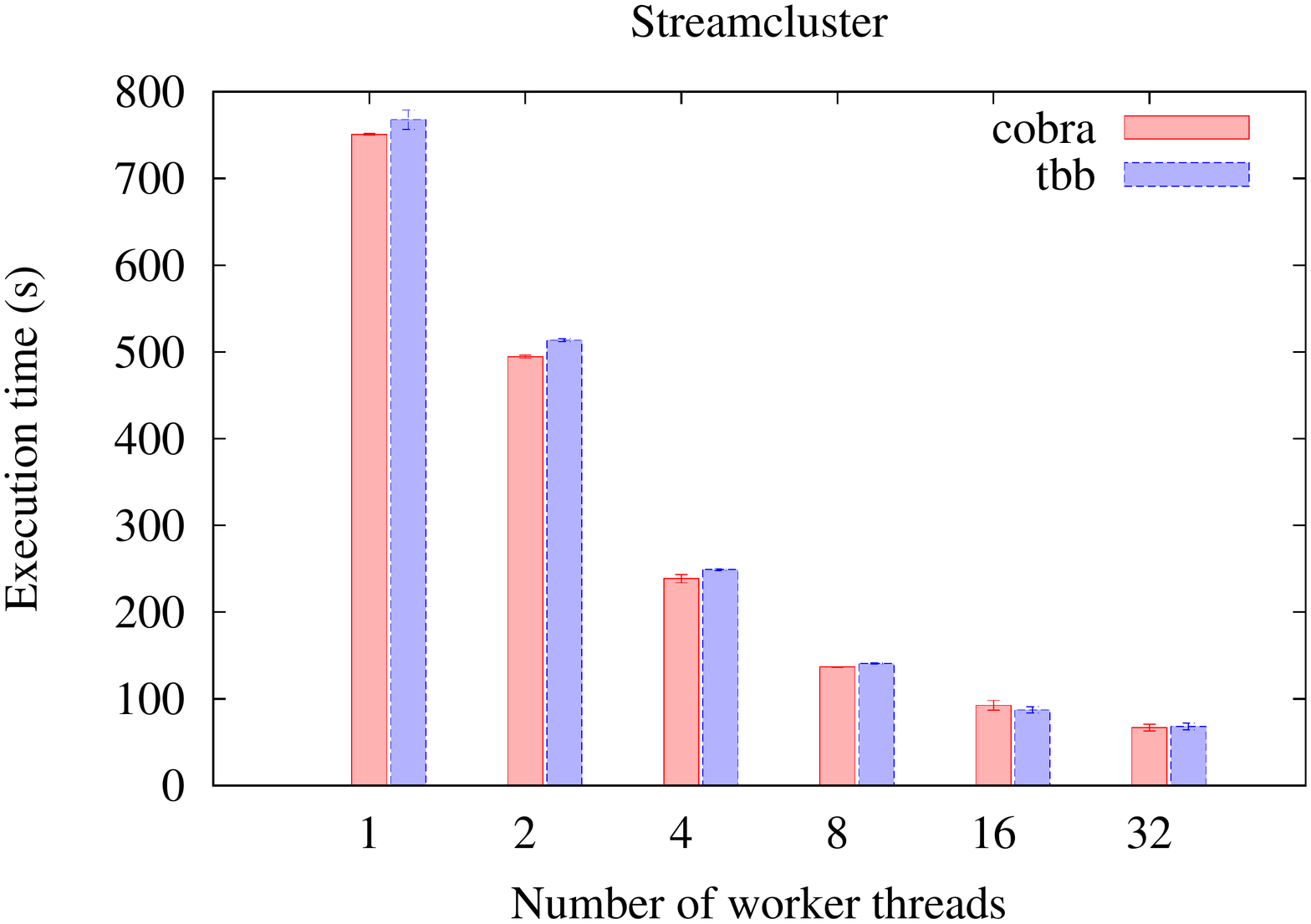}

&

\includegraphics[scale=0.185]{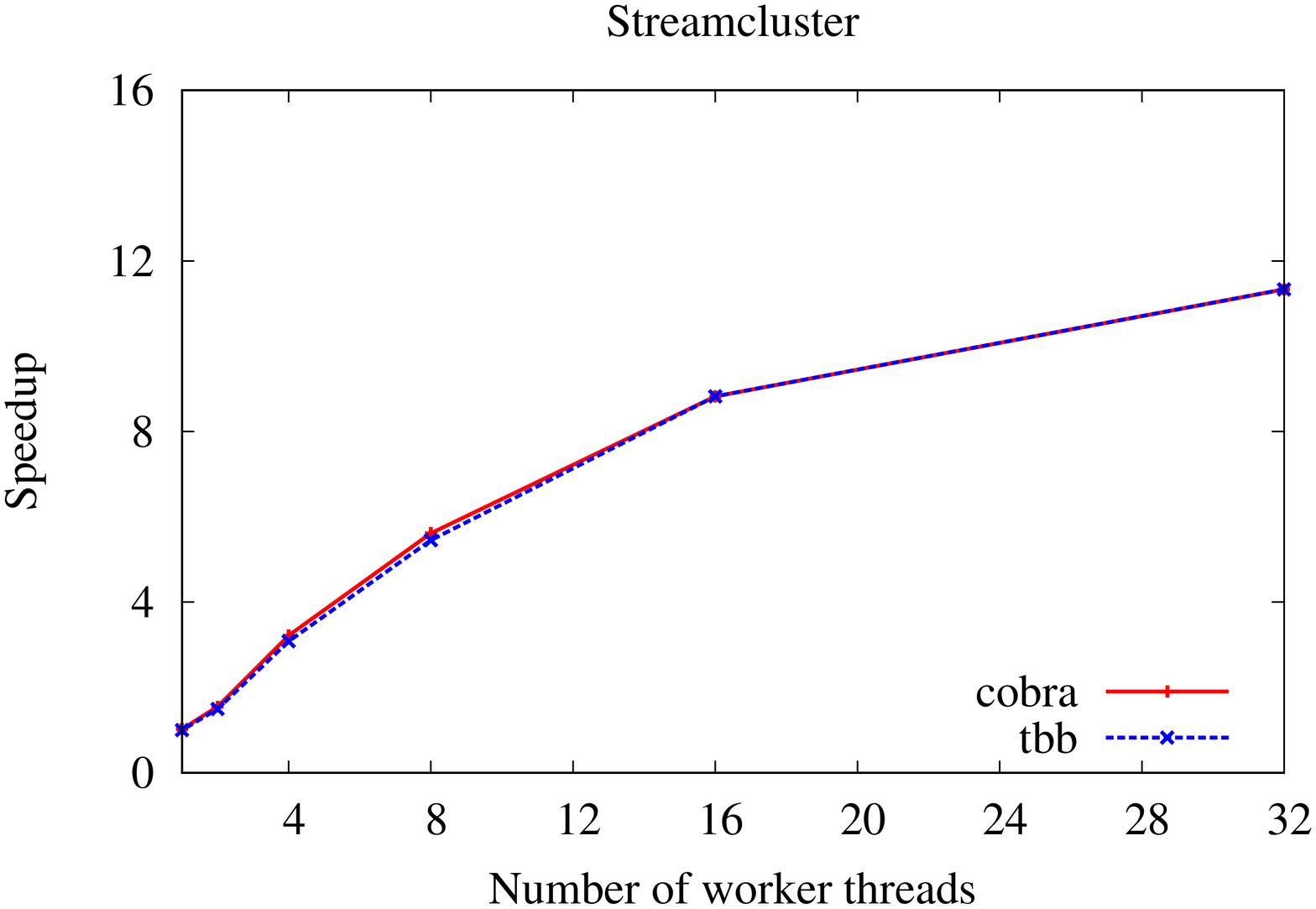}

&

\includegraphics[scale=0.185]{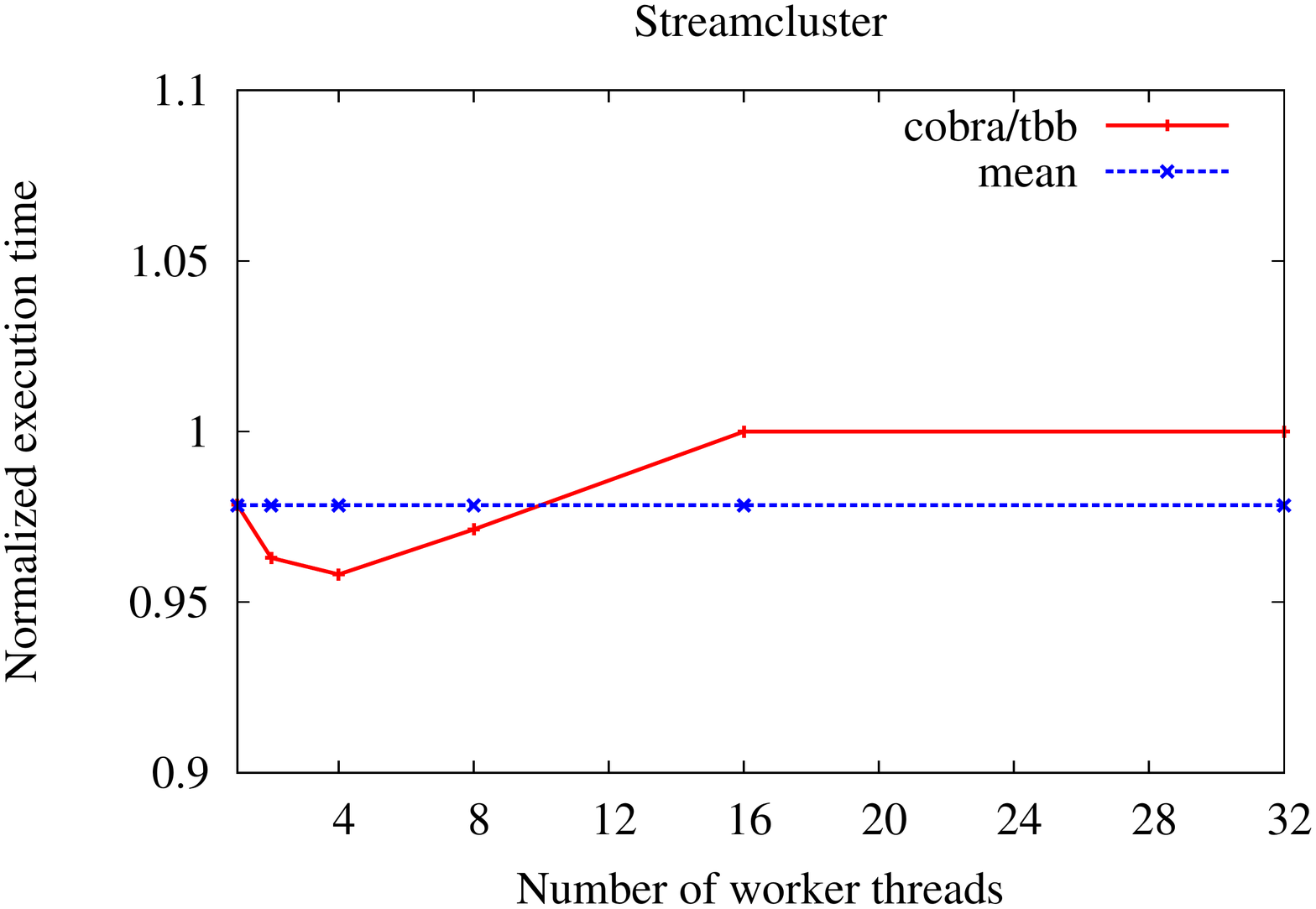}

\\

\includegraphics[scale=0.185]{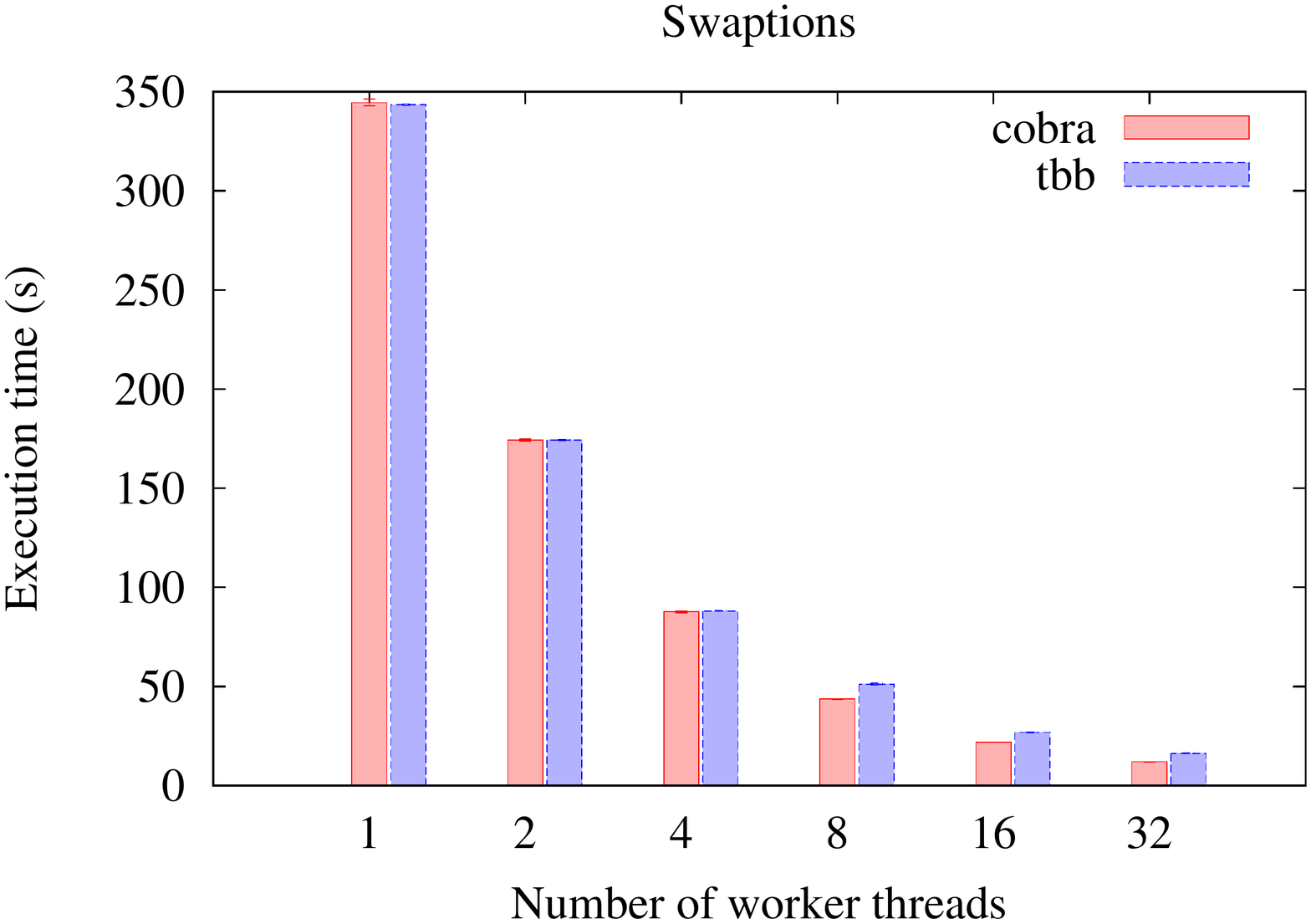}

&

\includegraphics[scale=0.185]{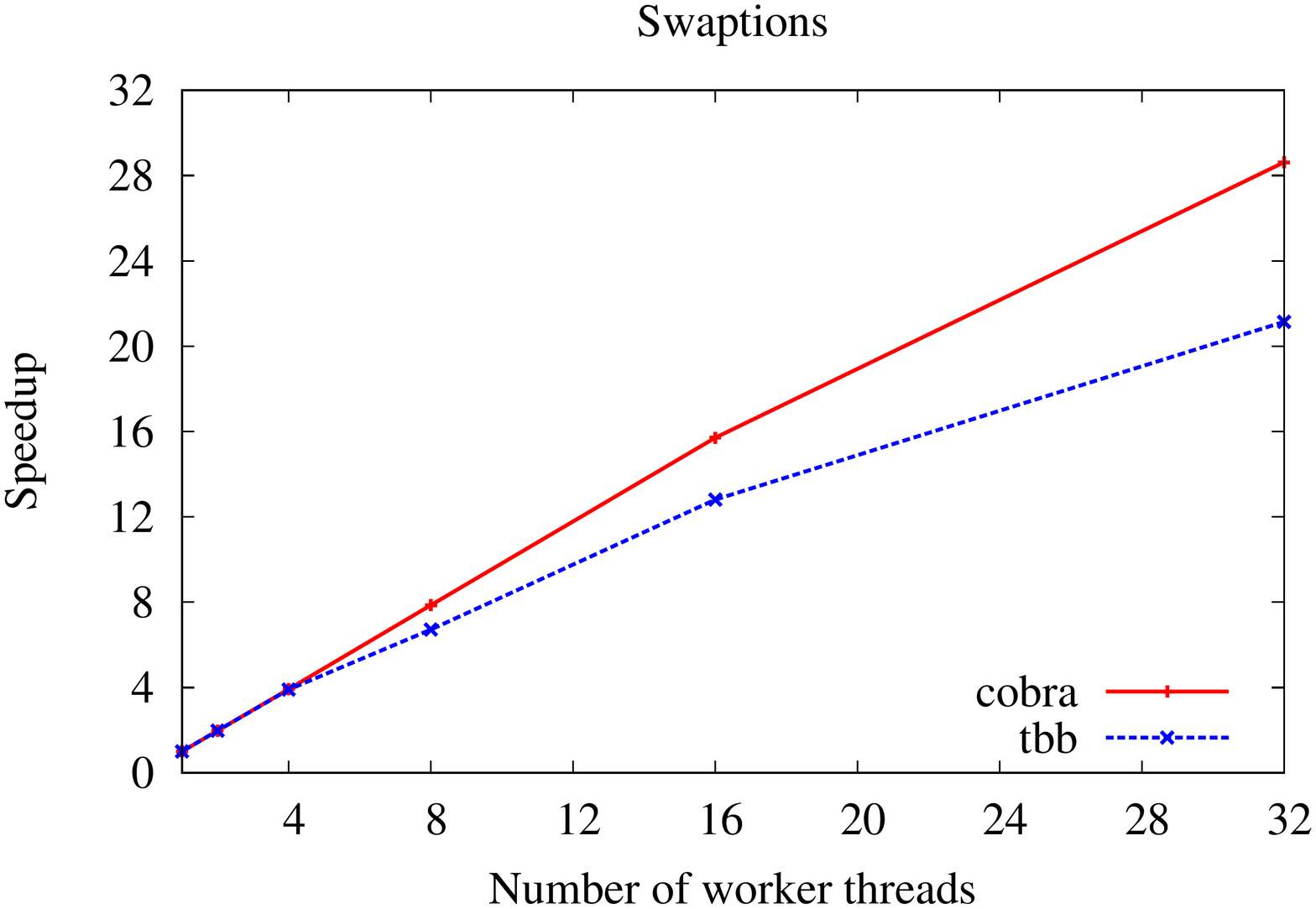}

&

\includegraphics[scale=0.185]{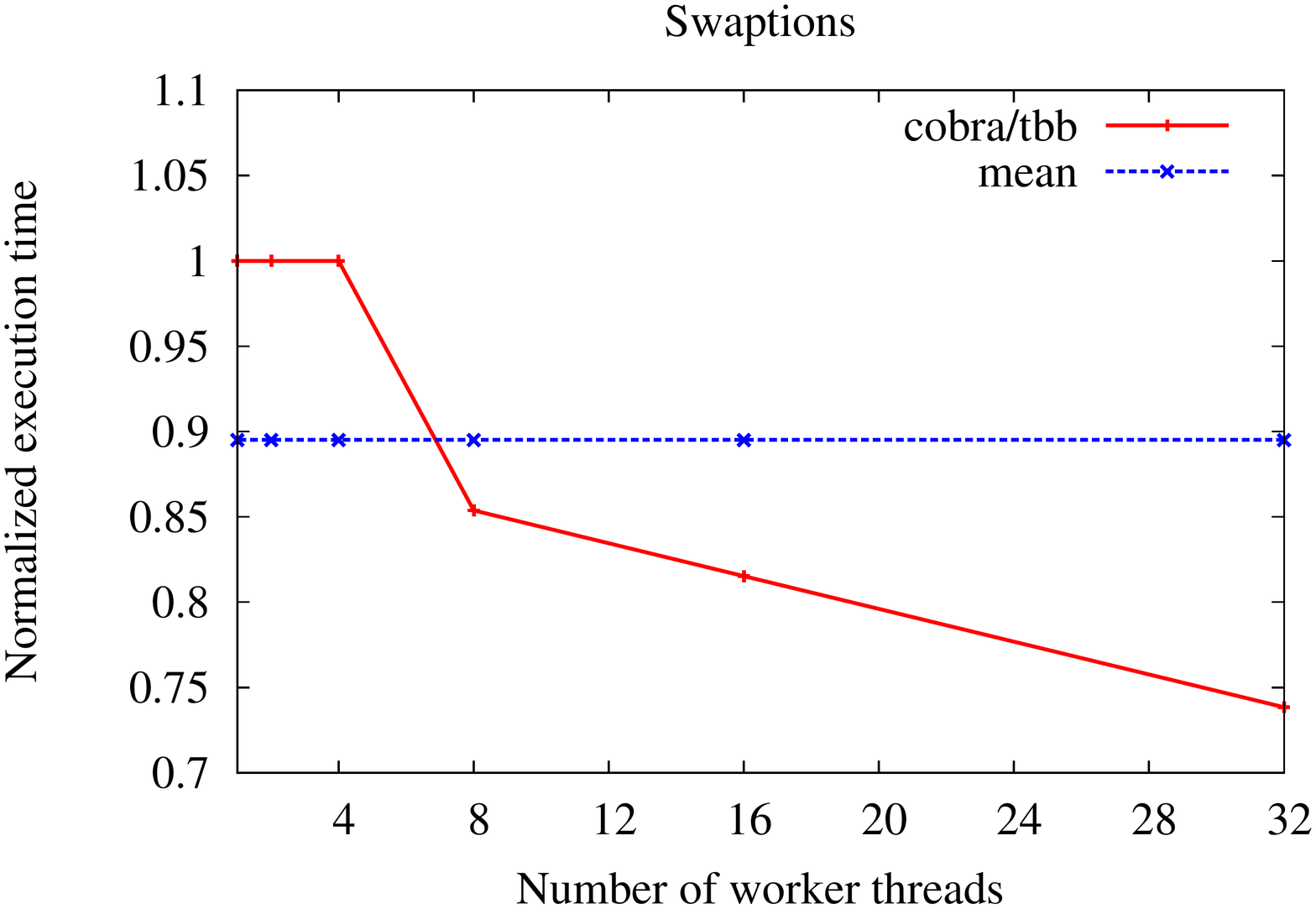}

\\

\end{tabular}
\caption{Cobra vs.\ TBB. 1) Raw execution times, showing the mean of $30$ runs with a $95\%$ confidence interval. 2) Speedup graphs for the mean of $30$ runs, using the execution on $1$ thread on TBB as the baseline. 3)
Relative performance
graphs depicting the mean execution time per thread on Cobra normalized to the mean execution times on TBB.
The blue lines depict the geometric means of the relative performance over all thread counts.
On average, Cobra is a factor $0.95$ away from TBB.}
\label{fig:overhead-resilience}
\end{figure*}

We compare the timings of the benchmarks on the work-stealing schedulers of both TBB and Cobra. Fig.~\ref{fig:overhead-resilience} shows the timings for respectively \textit{black\-scholes}, \textit{fluidanimate}, \textit{streamcluster}, and \textit{swaptions}. The figure shows three graphs per benchmark: a histogram comparing the raw execution times, a speedup graph depicting the scaling behavior, and a graph depicting the relative performance of Cobra compared to TBB. 

The histograms in Fig.~\ref{fig:overhead-resilience} compare the execution times of the benchmarks on Cobra (red) and TBB (blue) for $1$ to $32$ threads. Each bar represents the mean of $30$ runs. We also show the $95\%$ confidence intervals for each of those means. The confidence intervals are small, indicating a low variability between the timings of the different runs. Therefore, the speedup and 
relative performance graphs in Fig.~\ref{fig:overhead-resilience} are constructed using the means of the execution times. The confidence intervals also help identifying the statistically significant differences between runtimes on TBB and Cobra: Only the bars and intervals that do not overlap imply a statistically significant difference.\footnote{We also applied t-tests to verify the statistical significance.} When there is no statistically significant difference between the mean execution time on TBB and Cobra, the speedup and relative performance graphs in Fig.~\ref{fig:overhead-resilience} use the same means.

When comparing the timings of TBB and Cobra, we see that Cobra is typically on par with TBB for low thread counts ($1$ to $4$), but faster for higher thread counts ($8$ to $32$), except for the \textit{streamcluster} benchmark, where the execution times are roughly the same. This indicates that Cobra scales slightly better than TBB in certain cases, which is also confirmed by the speedup graphs in Fig.~\ref{fig:overhead-resilience}.

The third set of graphs in Fig.~\ref{fig:overhead-resilience} shows the (mean) execution times on Cobra normalized to the (mean) execution times on TBB (red). The normalized execution time gives a factor of how much slower/faster Cobra is compared to TBB. For thread counts from $1$ to $4$, we see that Cobra is between a factor $1$ and $1.06$ away from TBB. For higher thread counts, Cobra is between a factor $0.74$ and $0.99$ faster than TBB.

We can estimate the overall performance difference as a single number by computing the geometric mean of the relative performance factors~\cite{hennessy11}. These  means are also shown in the graphs with the blue lines. We get a mean of $0.97$ for \textit{blackscholes}, $0.97$ for \textit{fluidanimate}, $0.98$ for \textit{streamcluster}, and $0.90$ for \textit{swaptions}. By combining the timings for the different benchmarks, we get a mean of $0.95$, which means that, on average, Cobra has a similar performance as TBB.

We stress that the graphs in Fig.~\ref{fig:overhead-resilience} compare the \emph{unchanged} versions of the benchmarks on Cobra and TBB, although Cobra requires parts of the code to be idempotent to be able to recover from soft errors. We compare the unchanged versions of the benchmarks to focus on the relative performance of the resilience infrastructure, without measuring the impact of idempotence. For \textit{fluidanimate} and \textit{streamcluster}, which are not already idempotent in their unchanged versions, we measured that the idempotent variations of the code are, on average, a factor $1.08$ \emph{faster} than the non-idempotent versions, but this is not shown in the graphs. We believe that the idempotent code is faster because a) the idempotent coding style improves data locality, and b) the structure of the idempotent code can more easily be optimized by the compiler.

\subsection{Impact of failures}
\label{sec:impact-failures}

\begin{figure*}

\begin{tabular}{ccc}

\includegraphics[scale=0.185]{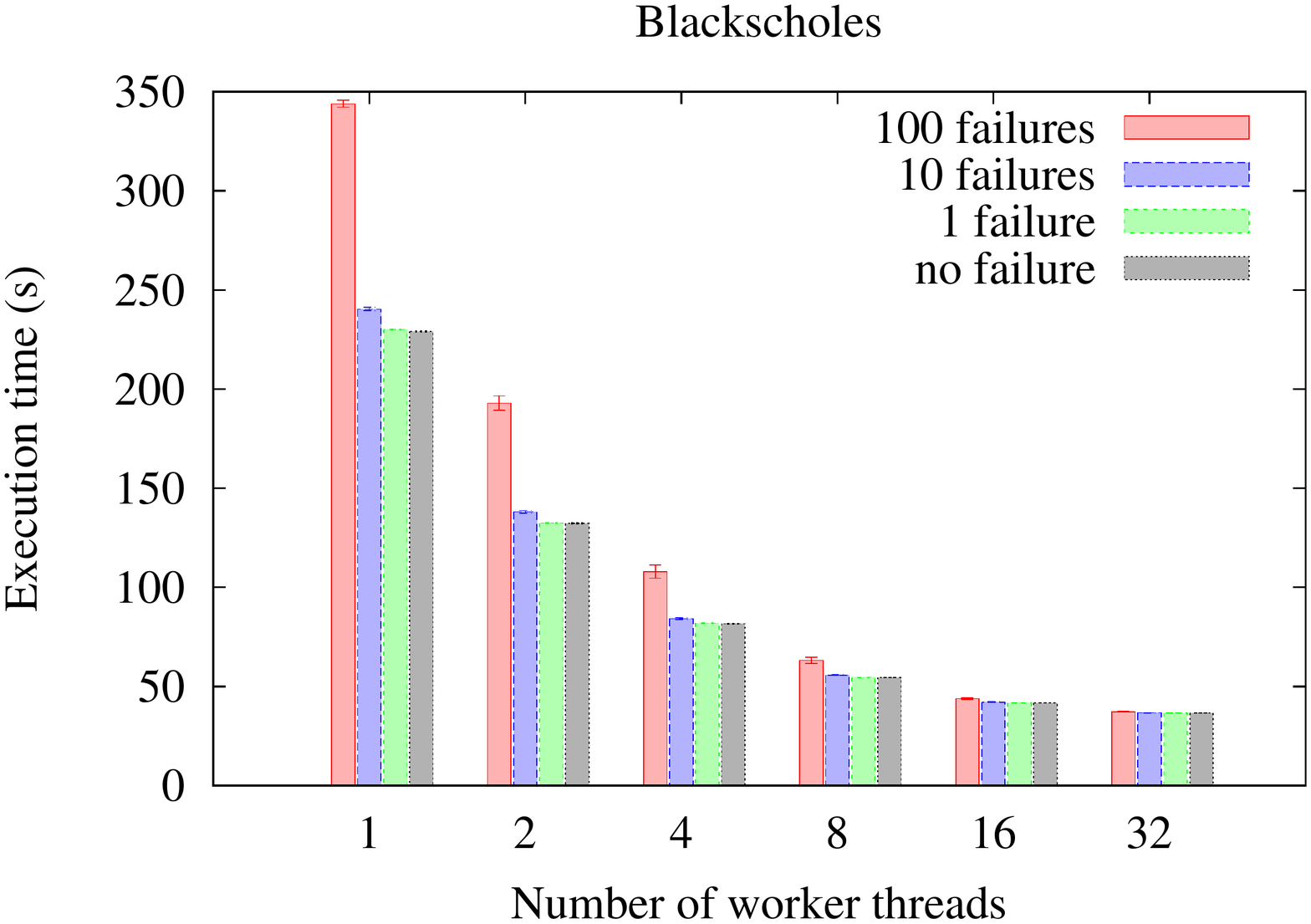}

&

\includegraphics[scale=0.185]{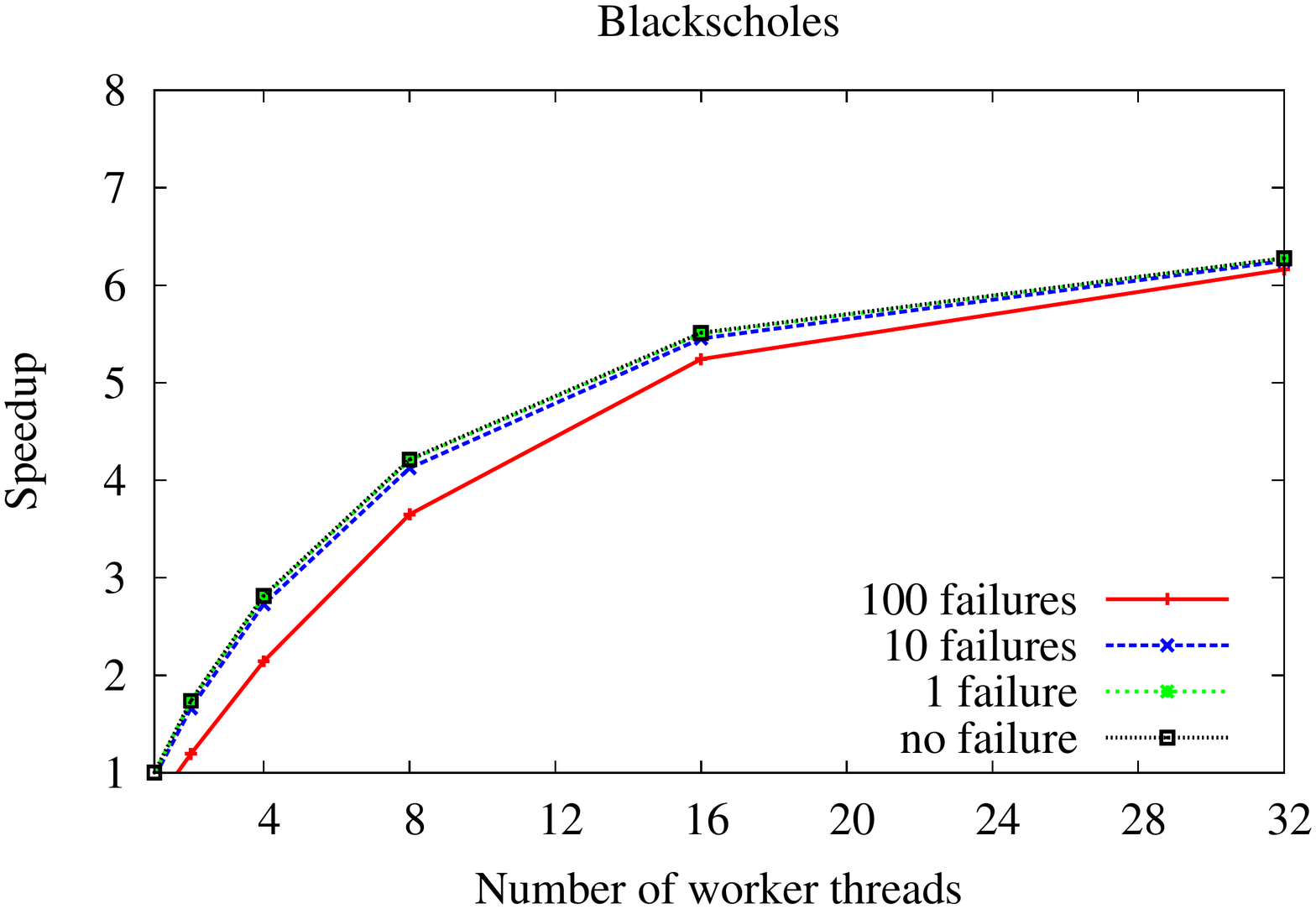}

&

\includegraphics[scale=0.185]{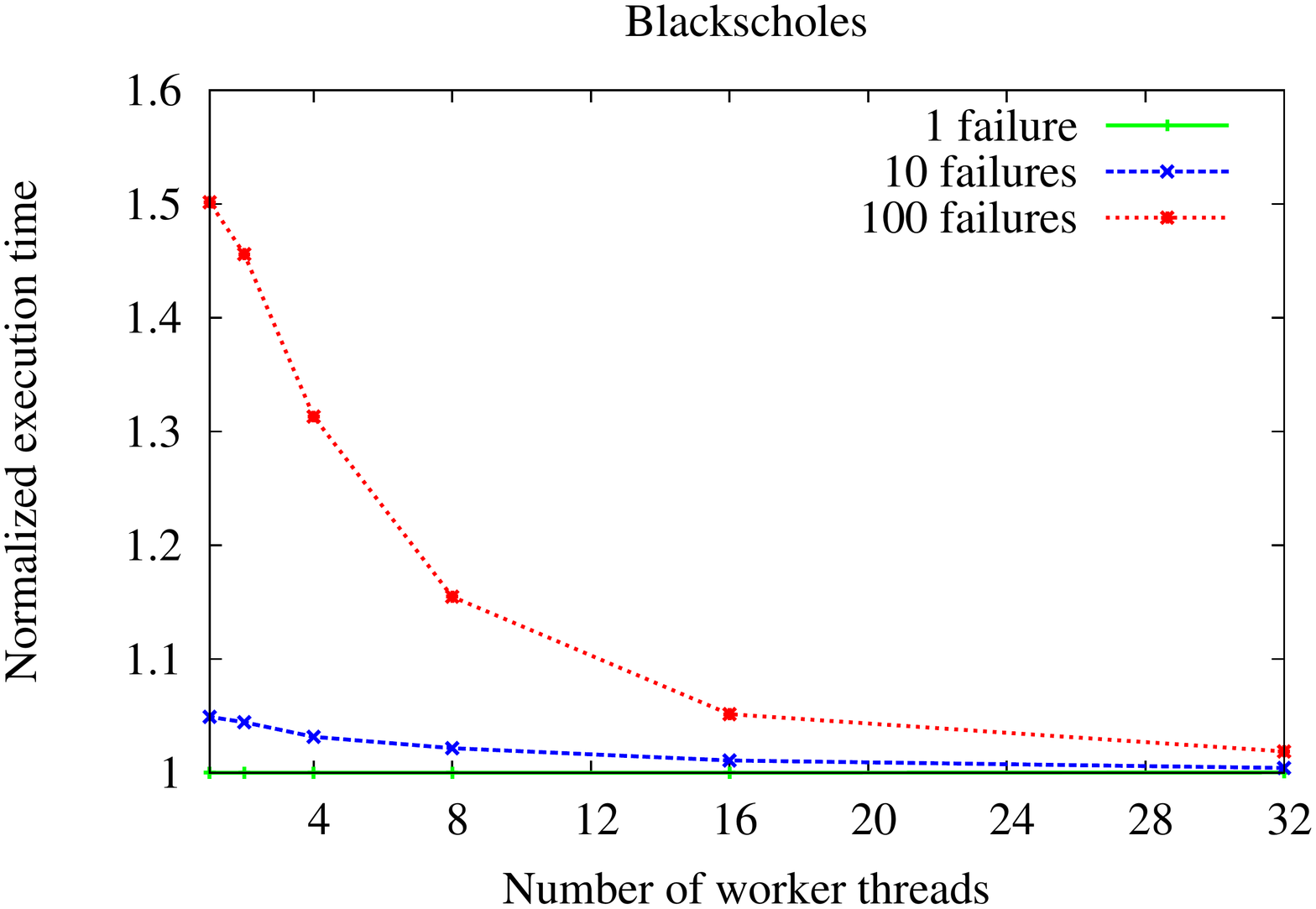}

\\

\includegraphics[scale=0.185]{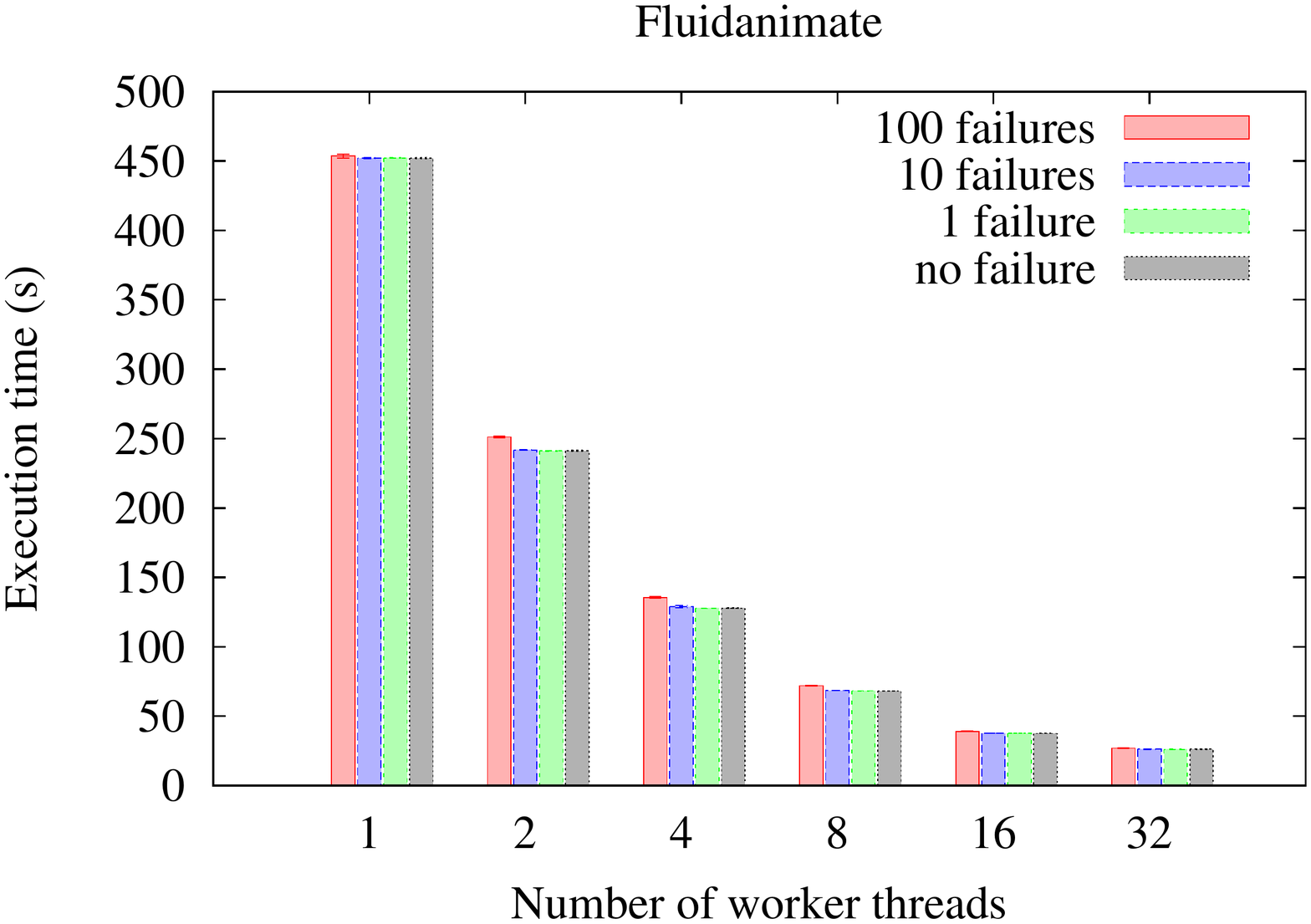}

&

\includegraphics[scale=0.185]{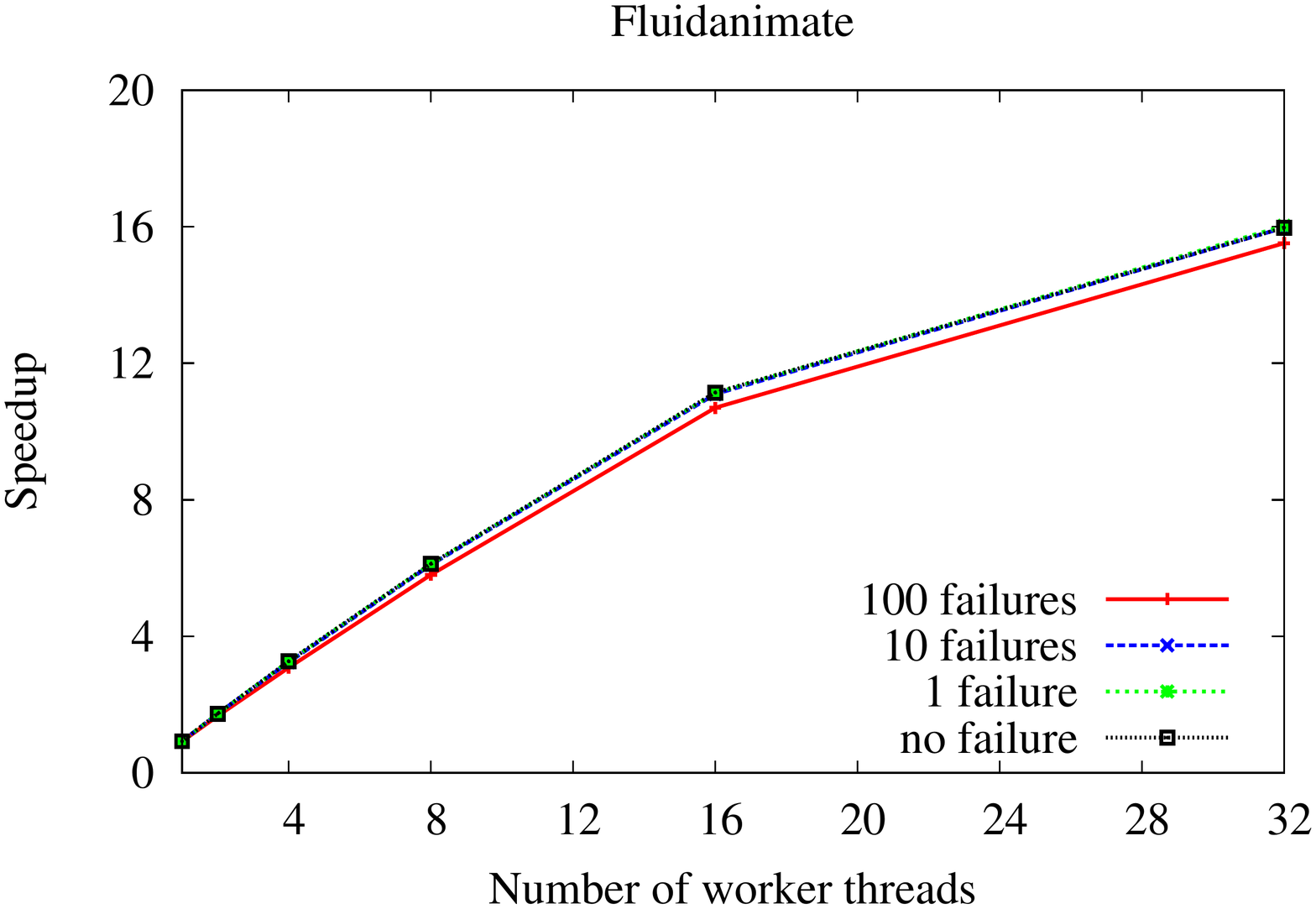}

&

\includegraphics[scale=0.185]{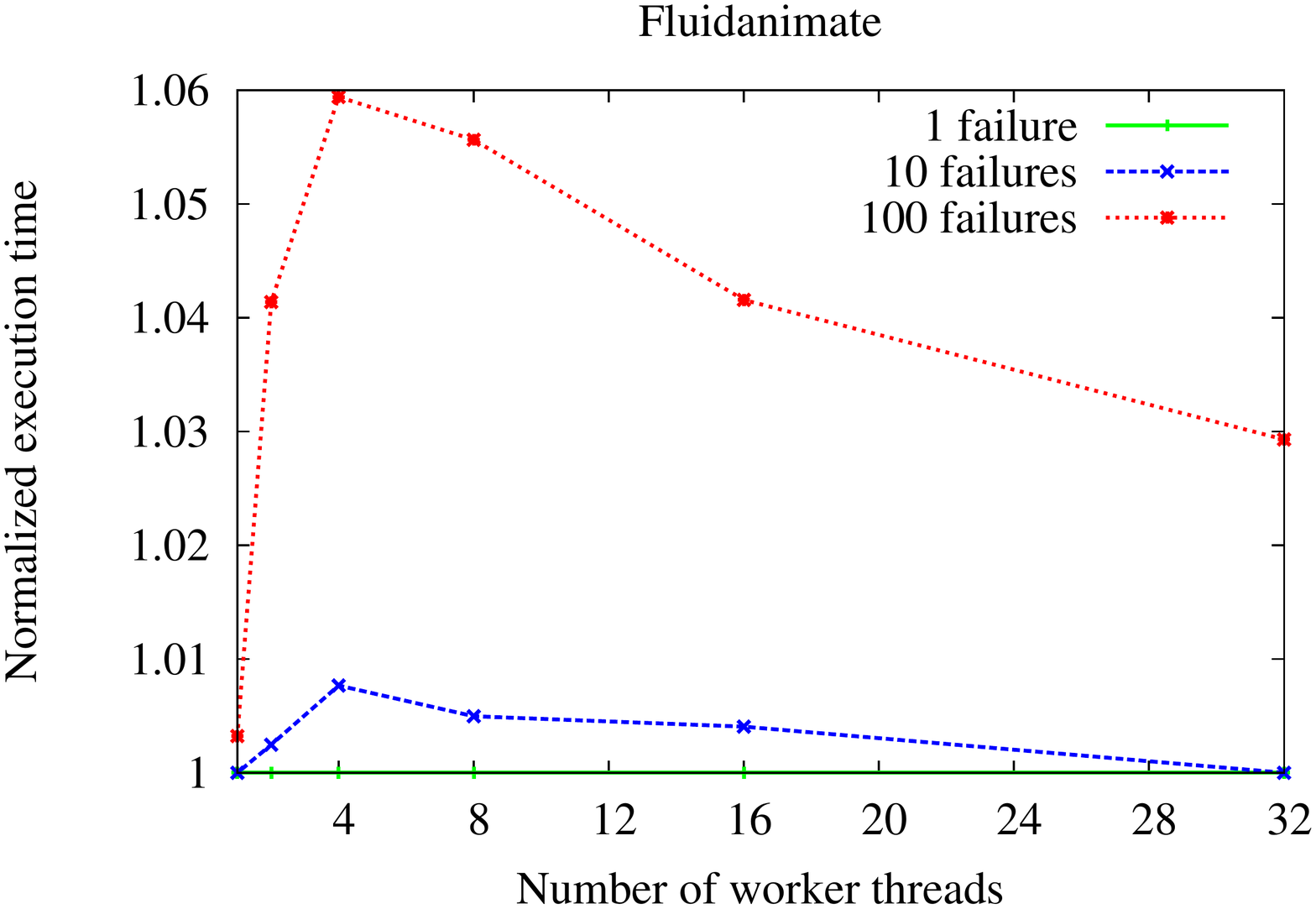}

\\

\includegraphics[scale=0.185]{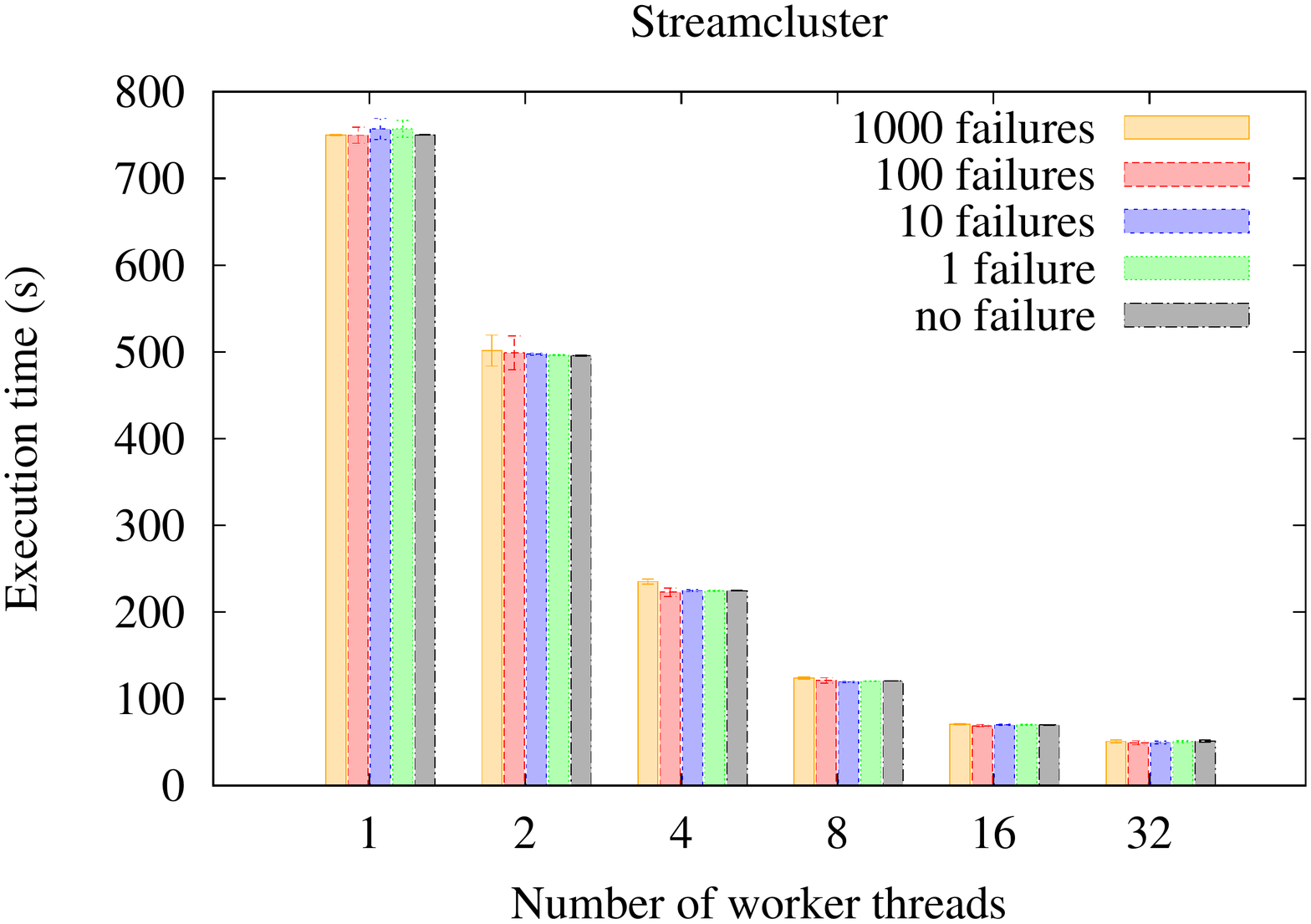}

&

\includegraphics[scale=0.185]{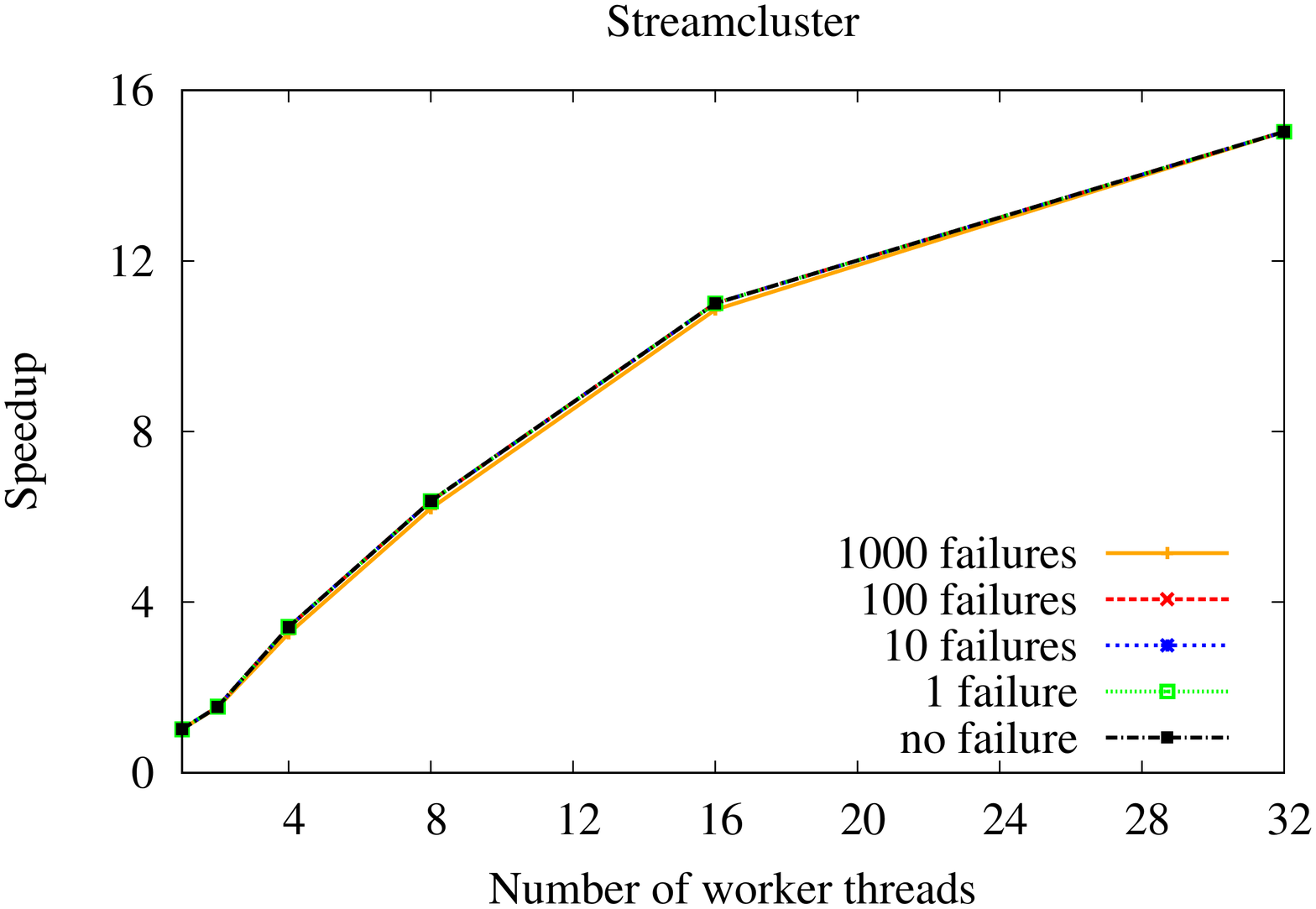}

&

\includegraphics[scale=0.185]{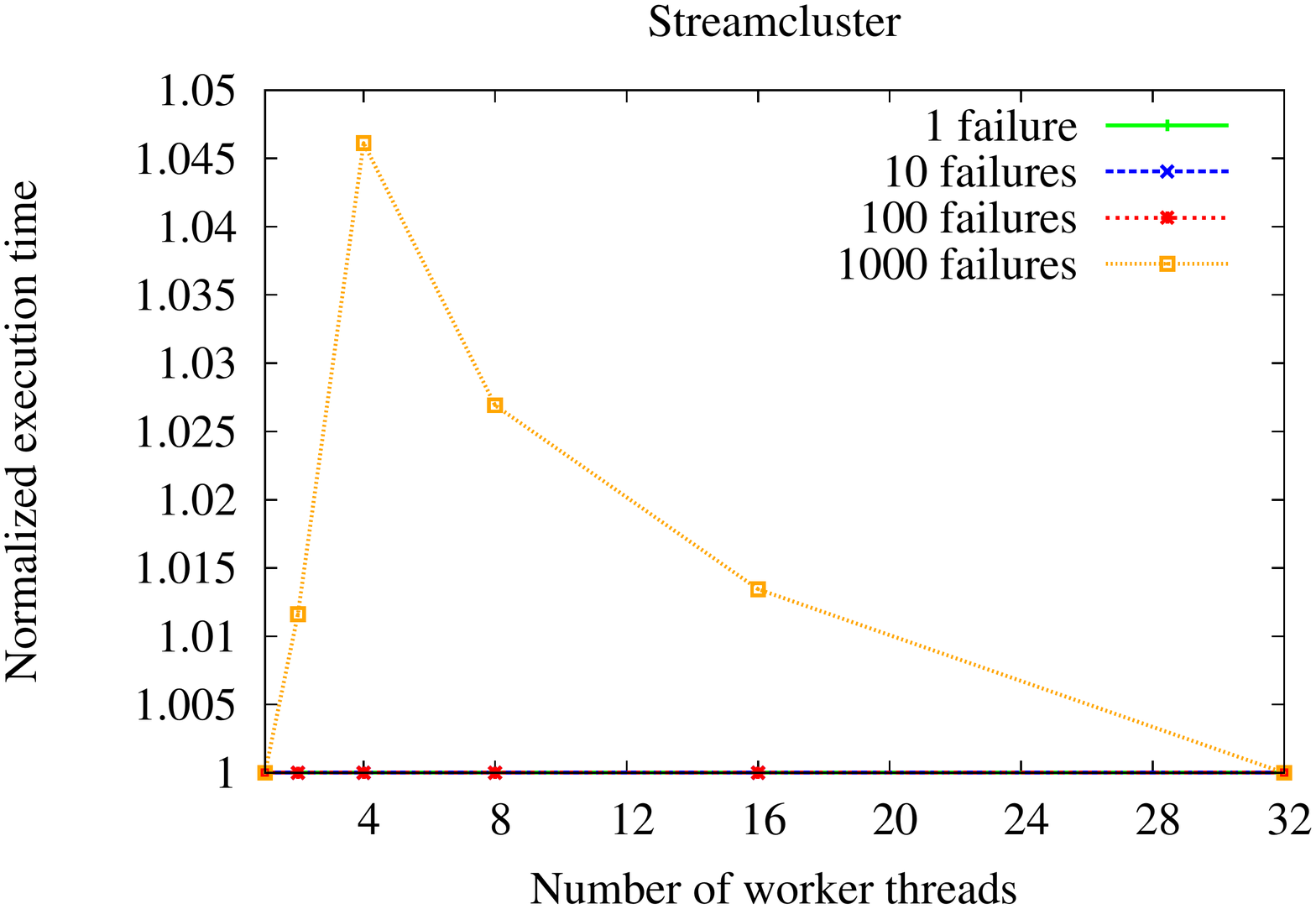}

\\

\includegraphics[scale=0.185]{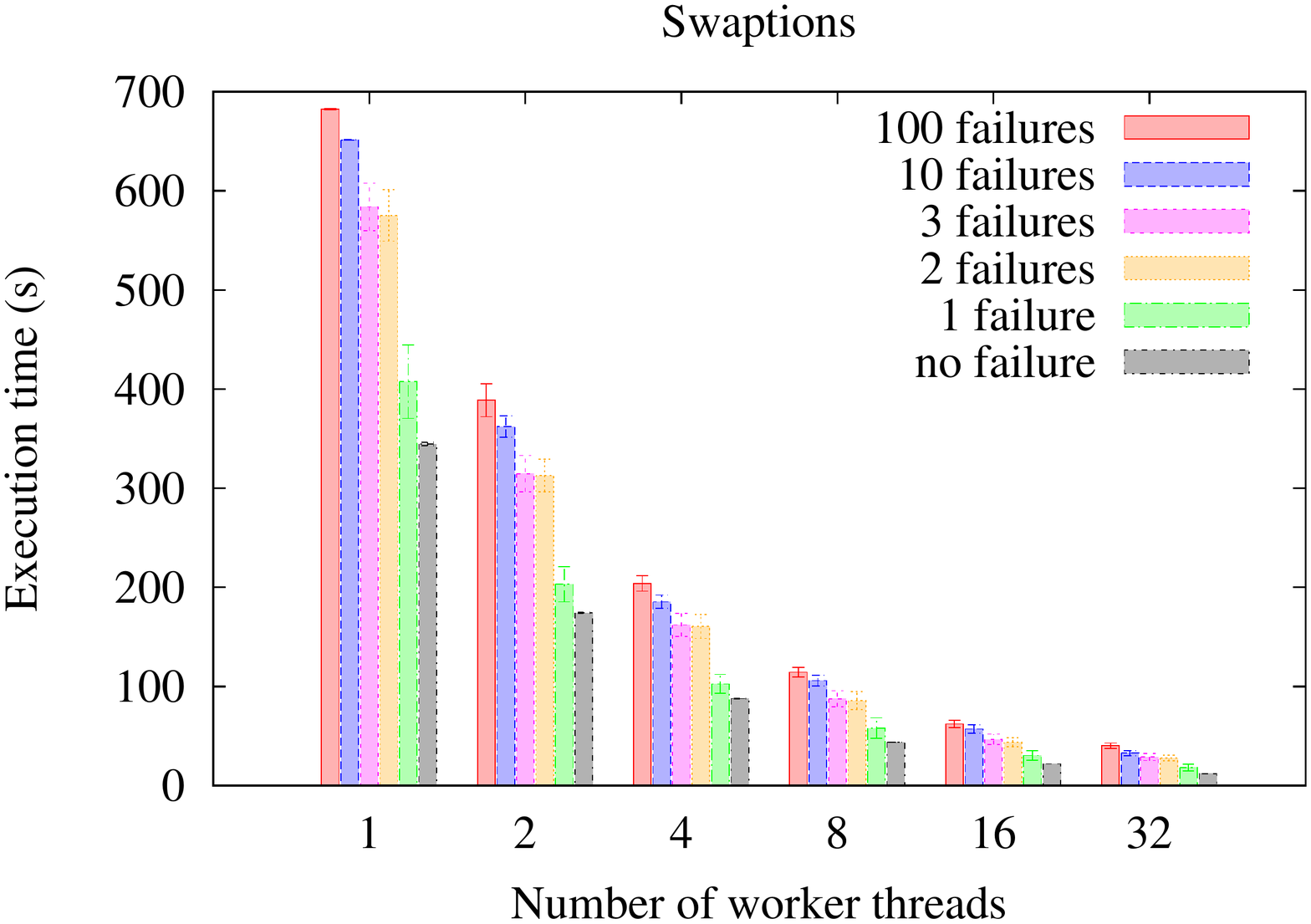}

&

\includegraphics[scale=0.185]{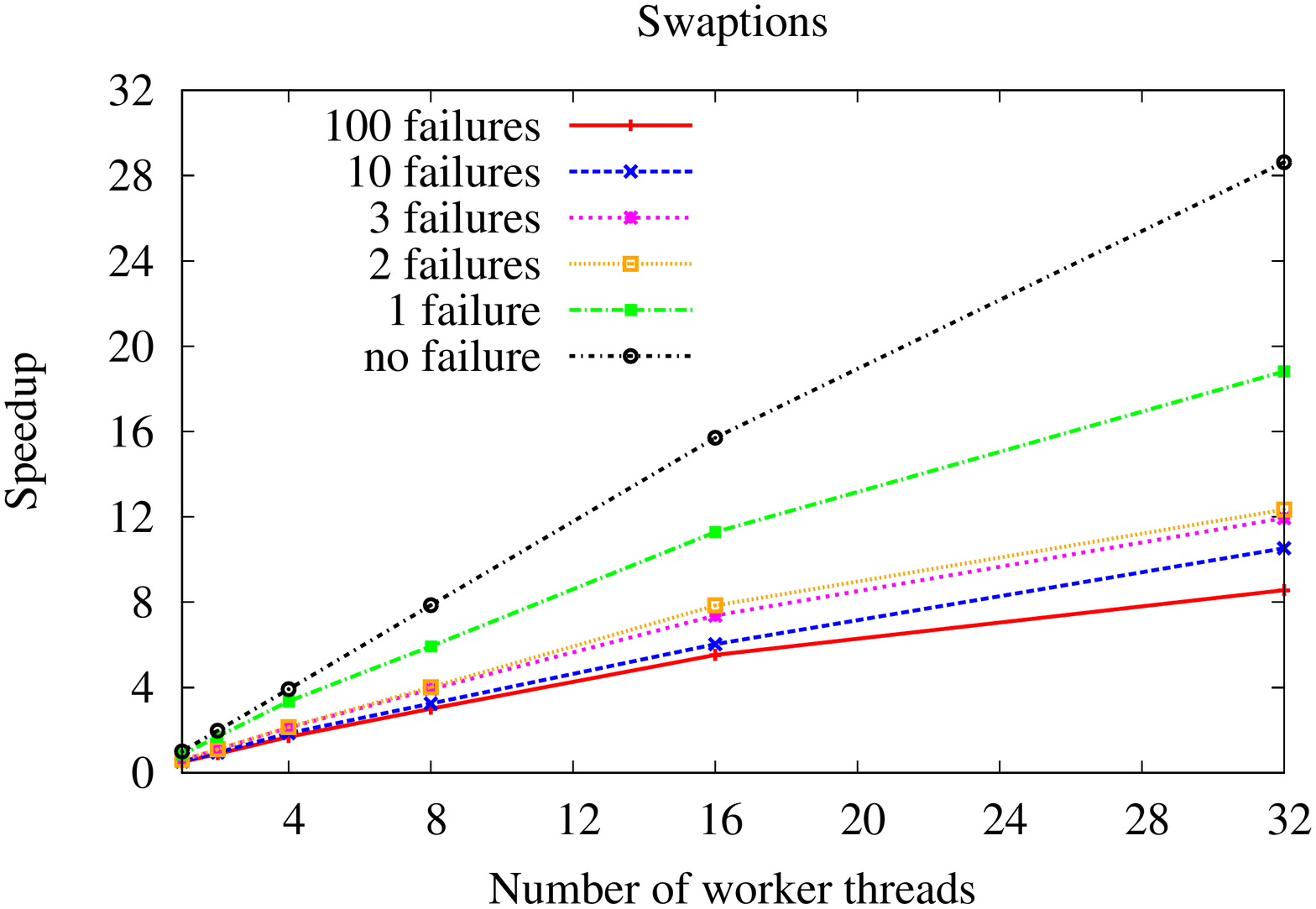}

&

\includegraphics[scale=0.185]{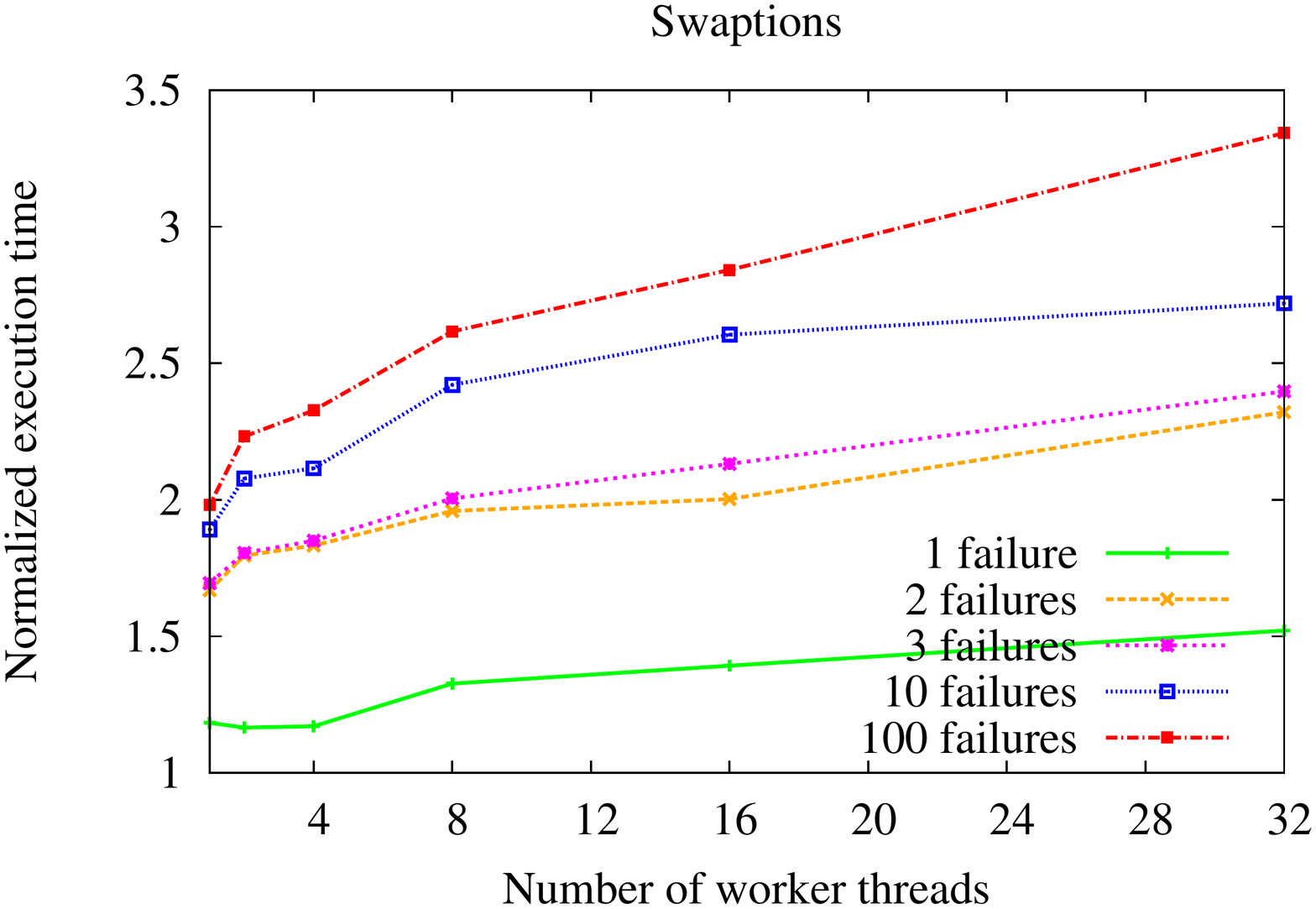}

\\

\end{tabular}
\label{fig:impact-crash}
\caption{Impact of $1$ to $1000$ failures per node. 1) Raw execution times, showing the mean of $60$ runs with a $95\%$ confidence interval. 2) Speedup graphs for the mean of $60$ runs, using the execution on $1$ thread on TBB as the baseline. 3) Overhead graphs depicting the mean execution time per thread on Cobra with $1$ to $1000$ (simulated) failures normalized to mean execution times on Cobra without failures. Although a conservative estimate is less than $10$ failures for the benchmarks' runtimes (Section~\ref{sec:failure-frequency}), we increased the failure rate until we got a statistically significant impact on the benchmarks' performance.}
\label{fig:impact-failure}
\end{figure*}

To investigate whether Cobra can survive hardware failures due to soft errors, and the performance when doing so, we set up a \textit{software-based fault injection} experiment.

\subsubsection{Experimental Setup}

The fault-injection experiment consists of 
a background thread that can asynchronously signal soft errors to worker threads using POSIX signals. This is similar to how the OS would signal soft errors to programs when encountering machine check exceptions~\cite{kleen09}. When a worker thread receives such a signal, it reinitializes its data structures and restarts its scheduling loop (Section~\ref{sec:design-monitoring}). 

Our simulation takes four parameters into account that determine the overall impact of a failure: the percolation degree, the target, the timing, and the frequency of failures. 

\subsubsection{Failure percolation} 
In Cobra, a node restart is caused by a read access to a poisoned memory location, which needs to be rewritten to become \textit{unpoisoned} again  (Section~\ref{sec:the-cobra-approach}).
In the best case, a restarted node immediately rewrites, and thus unpoisons that memory location. 
In the worst case, the restarting of nodes percolates up until the root node in the fork/join tree before the necessary rewrite is performed (Section~\ref{sec:design-monitoring}). To take this percolation into account, we do not simulate data poisoning directly, but instead simulate its effect on the computation. To simulate the worst-case effect, we modify the continuation of every restarted node such that it immediately triggers another failure when it is executed, which ensures percolation up to the root node. The root node's continuation is not modified, however, to ensure the overall computation still terminates. To simulate the best-case effect, no modification is necessary, since the computation can just proceed after a restart in this case.

\subsubsection{Failure target} 
Which target worker thread is selected for injecting a failure has an impact on the overall performance. 
Injecting a failure in a worker thread that is in its steal loop, because it happens to currently have no work of its own, has a 
lower impact than selecting a worker thread that is busy with nodes deep down the fork/join tree. In work stealing, 
which worker thread works on what part of a fork/join tree is randomized. In our failure simulation, we therefore also randomly 
choose the worker threads that get failure signals. To get a fair selection of worker threads that are failed, we run each 
failure experiment thirty times and average the timings.

\subsubsection{Failure Timing} 
The timing of a failure may have an effect on performance. To guarantee a fair selection of failure times, our simulation schedules failures over the average execution time without failures, which is measured for each benchmark and each configuration upfront.

\subsubsection{Failure frequency} 
\label{sec:failure-frequency}
 The frequency of failures predicted in the literature varies among different sources, but can be estimated using the architectural vulnerability factor (AVF)~\cite{mukherjee08}. A rough estimate is that future processors will have a mean-time-to-failure (MTTF) of months, but for clusters which are composed of thousands of processors this is expected to result in a MTTF of minutes~\cite{borkar05,cappello09,debardeleben09,dongarra11,elnozahy08,geist11,rivers09,schroeder07}. 
Our benchmarks are also in the range of minutes, so a conservative estimate is that we can expect between $1$ and $10$ failures. We choose to inject at least between $1$ and $100$ failures per experiment, and for one benchmark, we did experiments with $1000$ failures. 

\subsubsection{Results} 

Cobra survives the failures injected in our experiments: It runs the benchmarks to completion and still produces correct results in the presence of failures, confirmed by validating their output to failure-free execution of the unchanged equivalent programs running on TBB. 

In Fig.~\ref{fig:impact-failure}, we show the impact on the performance of our benchmarks when injecting failures according to the above strategy. Three graphs are depicted for each of our benchmarks: a histogram for comparing the execution time on Cobra with $1$ to $1000$ failures to the execution time on Cobra without failures, a speedup graph to illustrate the impact of failures on the scaling behavior, and an overhead graph showing the execution time with failures normalized to the failure-free execution time.

The histograms for comparing execution times depict the mean of $60$ runs with a $95\%$ confidence interval. The execution times without failures are those for the idempotent versions of the benchmarks. The confidence intervals tell us where differences with failure-free execution are statistically relevant, and we use this information for the speedup and overhead graphs in Fig.~\ref{fig:impact-failure}.\footnote{We also applied t-tests to verify the significance.}

In general, as the amount of failures increases, the execution gets slower. The impact, however,  is highly benchmark-specific. For \textit{swaptions}, we already see statistically significant differences with failure-free execution starting from runs with $1$ failure. In contrast, for  \textit{blackscholes} and \textit{fluidanimate}, it requires at least $10$ failures to see a (small) significant impact. For \textit{streamcluster} it takes even $1000$ failures before there is any observable impact. These differences are due the differences in granularity of the fork/join computations in the benchmarks. When a failure occurs and a worker thread restarts, then in the worst case, all fork/join computations that are (partially) worked on by that worker thread need to be completely restarted. So if a program generates many short small fork/join trees, then the impact of a failure will be lower than when a program generates fewer, but larger and longer-lasting fork/join trees. For example, \textit{swaptions} generates exactly 1 fork/join tree in each run, hence the impact of a single failure is high ($1.2$ to $1.5\times$ slower), whereas \textit{streamcluster} generates $53460$ fork/join trees and the impact of $1000$ failures is low ($1.01$ to $1.05\times$ slower). Table~\ref{tab:average-impact-failures} lists the number of fork/join trees per benchmark, as well as the average slowdown per number of failures obtained by calculating the geometric mean of the overhead numbers in Fig.~\ref{fig:impact-failure}.

\begin{table}
\begin{center}
\begin{footnotesize}
\begin{tabular}{|l|l|l|l|l|l|}
\hline
Benchmark & \#Trees & \multicolumn{4}{c|}{\#Failures} \\
\hline
 & & 1 &  10 & 100 & 1000 \\
\hline
Swaptions & 1 & 1.29 & 2.29 & 2.52 & n/a \\
\hline
Blackscholes & 100 & 1.00 & 1.03 & 1.24 & n/a \\
\hline
Fluidanimate & 3500 & 1.00 & 1.01 & 1.05 & n/a \\
\hline
Streamcluster & 53460 & 1.00 & 1.00 & 1.00 & 1.02 \\
\hline
\end{tabular}
\end{footnotesize}
\end{center}
\caption{Average slowdown per benchmark per number of failures, related to the granularity of the fork/join computations.}
\label{tab:average-impact-failures}
\end{table}

To get an idea of how well Cobra handles failures, we compare the numbers in Table~\ref{tab:average-impact-failures} to the slowdowns we can expect with a pure checkpoint-restart system. Consider the case for 1 failure. Assume the execution time without failures is $T$. In a pure checkpoint-restart approach, the execution time with 1 failure is $T + \alpha T$, with $\alpha$ a factor that determines how far the execution progressed before the failure occurred.\footnote{Assuming the restart time itself is $0$, which is optimistic.} In the best case, the failure occurs at the beginning and $\alpha=0$, so the total execution time is $T$, as if no failure occurred. In the worst case, the failure occurs just before the program finishes and $\alpha=1$, so the total execution time is $2T$. However, we have to assume that the failure occurs with equal chance for some point in the interval $[0,T]$, so on average $\alpha=0.5$ and the average execution time with 1 failure is $1.5T$. With Cobra, the expected slowdown for 1 failure is a factor $1.00$ to $1.29$, as shown in Table~\ref{tab:average-impact-failures}. This is well below the $1.5$ slowdown we estimate for a pure checkpoint-restart approach. 

The above reasoning can be generalized to executions with multiple failures. Assume that, on average, the failures are evenly distributed over the time interval $[0,T]$. Hence for a checkpoint-restart approach, the execution time for $n$ failures is $\sum_{\alpha=1}^{n+1} \frac{\alpha}{n+1}T = (1+\frac{n}{2})T$. So for $10$ failures, the expected execution time is $6T$, for $100$ failures it is $51T$, and for $1000$ failures it is $501T$. Table~\ref{tab:average-impact-failures} shows that the slowdowns measured with Cobra are far below these estimates.

\section{Related work}

\subsection{Software resilience}

In the field of high-performance computing, the interest in fault tolerance at the software level has recently increased due to higher software-visible failure rates being reported for real-world clusters \cite{debardeleben09,schroeder07,geist11}. Recent papers classify existing approaches according to whether they mask failures or not, and whether they avoid failures altogether, avoid the effects of failures, or repair such effects \cite{cappello09b, cappello09, debardeleben09}.

\emph{Failure-avoiding strategies} move computations away from hardware components before they are predicted to fail. Such strategies may suffer from false positives and false negatives, and need to be complemented by other fault tolerance approaches that deal with unpredicted failures \cite{cappello09b}.

\emph{Failure-effect avoiding strategies} include, for example, replication, ABFT, and naturally fault-tolerant algorithms. Replication performs all computations more than once, such that a failure can be masked by the successful computations. ABFT augments data structures with checksums, and computations on the data structures modify such checksums accordingly. Results can then be double-checked after a computation finishes, and corrected if necessary. Naturally fault-tolerant algorithms are designed to handle silent errors. For example, certain classes of iterative linear solvers can be designed such that erroneous values in intermediate results delay, but do not prevent convergence to the correct result. ABFT and naturally-fault tolerant algorithms need to be explicitly designed, and thus can be classified as non-masking approaches \cite{cappello09}.

\emph{Failure-effect repair} is typically achieved by rollback recovery, with checkpointing being the most widely used fault tolerance approach, since it has little performance overhead for failure-free execution, and requires little involvement from  applications. Writing checkpoints to secondary storage incurs a cost, however, and different variations of checkpointing yield different costs at different scales \cite{bosilca12}. Due to these costs, checkpointing is predicted to become infeasible in the future if not augmented by other techniques. There are both masking and non-masking variants, depending on the necessary involvement from applications. Forward recovery is another, non-masking technique for failure-effect repair in which applications explicitly reconstruct correct computational states after failure \cite{cappello09}.

According to these criteria, our notion of restartable task graphs can be classified as a failure-effect repair approach. Cobra currently requires computations to be idempotent to ensure resilience. Except for this restriction, Cobra does not require any awareness of failures at the application level, and thus can be regarded as a masking approach.

\subsection{Work-stealing schedulers}

In existing work-stealing schedulers, worker threads maintain double-ended queues (deques), where work can be locally pushed and popped on one end, and remotely stolen from the other end. Work-first and help-first variations of this approach differ in what part of a computation is pushed on the thread-local deque by a fork, and which threads will eventually execute the continuations of suspended joins \cite{guo09}. 

\pagebreak

Our experiments to add resilience to exising approaches showed that work-first schedulers make it difficult to determine which threads need to check which parts of the fork/join tree, and which suspended joins need to be dropped in the presence of failures. On the other hand, help-first schedulers traditionally use the native call stack for maintaining continuations, and may therefore induce a long delay before noticing a failed subcomputation when it is ``hidden'' by other computations deeper down in the same call stack. Cobra is a variation of a help-first scheduler that, however, uses an explicit representation of the fork/join tree, which allows for checking any of its dependencies at any time.

In current work-stealing schedulers, a successful attempt at stealing work implies that it is immediately removed from the victim deque. In Cobra, stolen tasks are not immediately removed, but only flagged, so that parent nodes can check the liveness of child nodes. On top of that, existing work-stealing schedulers use fork counters to determine how many tasks are unfinished. Such fork counters are inherently non-idempotent, and thus Cobra avoids them to ensure that the scheduler code itself is resilient as well.

\section{Conclusions}
\label{sec:conclusions}

To deal with the increasing reliability issues at the hardware level, we propose to model computations in terms of \emph{restartable task graphs}, so that failures can be handled in software. As a proof of concept, we present \emph{Cobra}, a novel design for shared-memory work stealing that implements the notion of restartable task graphs and makes it possible for computations to recover from \emph{soft errors}.

In Cobra, the scheduler maintains an \emph{explicit representation of tasks} and tracks the dependencies between tasks. This allows for restarting any task whose execution is interrupted by a failure. As a \emph{fault detection mechanism}, we rely on the \emph{machine check architecture}, a modern processor design that can signal software-recoverable failures to the operating system. The Cobra scheduler responds to such \emph{machine check exceptions} triggered upon detecting \emph{soft errors} by searching for the task that rewrites the \emph{poisoned} memory location. This task cannot be determined upfront, and thus the scheduler starts by re-executing the task that was initially  interrupted. If this task does not override the poisoned memory, another machine check exception is triggered, and the scheduler attempts to restart a parent of the task instead. This repeats until, eventually, a task is re-executed that fixes the cause of the failure. This \emph{restart mechanism for tasks} is integrated as part of the regular scheduling loop, and therefore incurs little performance overhead.

As a validation, we executed the programs of the PARSEC benchmark suite that are based on Threading Building Blocks (TBB) on top of a C++11 implementation of Cobra. Comparison of failure-free execution with the work-stealing scheduler of TBB shows that Cobra is, on average, a factor $0.95$ away from TBB. This indicates that the performance of both schedulers is very similar. We also set up a \emph{software-based fault-injection} experiment. Our experiments confirm that Cobra executes programs correctly both with and without failure injection. We also observe that injecting failures yields an acceptable slowdown. For 1 failure we see a slowdown of $1.00$ to $1.29$ and for 100 failures a slowdown of $1.00$ to $2.52$, which is only a fraction of what can be expected with a conventional checkpoint-restart approach.


\begin{small}

\end{small}


\begin{thebibliography}{}

\bibitem{amd12} BIOS and Kernel Developer's Guide for AMD Family 15h Models 00h-0Fh Processors, Advanced Micro Devices, 2012.

\bibitem{bienia08} C. Bienia, S. Kumar, J.P. Singh, K. Li, The PARSEC Benchmark Suite: Characterization and Architectural Implications, PACT'08, ACM Press.

\bibitem{blumofe99} R.D. Blumofe, C.E. Leiserson, Scheduling Multithreaded Computations by Work Stealing, Journal of the ACM, 9, 1999.

\bibitem{borkar05} S. Borkar, Designing Reliable Systems from Unreliable Components: The Challenges of Transistor Variability and Degradation, IEEE Micro, November/December 2005.

\bibitem{bosilca12} G. Bosilca, A. Bouteiller, E. Brunet, F. Capello., J. Dongarra, A. Guermouche, Y. Herault, F. Vivien,  D. Zaidouni, Unified Model for Assessing Checkpointing Protocols at Extreme-Scale, INRIA technical research report no. 7950, May 2012.

\bibitem{cappello09} F. Cappello, A. Geist, B. Gropp, S. Kale, B. Kramer, M. Snir, Toward Exascale Resilience, Technical Report INRIA-Illinois Joint Lab on PetaScale Computing, TR-JLPC-09-01, 2009.

\bibitem{cappello09b} F. Capello, Fault Tolerance in Petascale/Exascale Systems: Current Knowledge, Challenges and Research Oppurtunities, The International Journal on High Performance Computing Applications, 23(3), 2009.

\bibitem{charles05} P. Charles, C. Grothoff, V.A. Saraswat, C. Donawa, A. Kielstra, K. Ebcioglu, C. von Praun, V. Sarkar, X10: An Object-Oriented Approach to Non-Uniform Cluster Computing, OOPSLA'05, ACM Press.

\bibitem{debardeleben09} N. DeBardeleben, J. Daly, S. Scott, C. Engelmann, B. Harrod, High-End Computing Resilience: Analysis of Issues Facing the HEC Community and Path-Forward for Research and Development, December 2009.

\bibitem{dixit11} A. Dixit, A. Wood, The impact of new technology on soft error rates, IRPS'11, IEEE, 2011.

\bibitem{dongarra11} J. Dongarra, P. Beckman et al., The International Exascale Software Project Roadmap, International Journal of High Performance Computer Applications, 25(1), 2011.

\bibitem{duran08} A. Duran, J. Corbal\'{a}n, E. Ayguad\'{e}, Evaluation of OpenMP Task Scheduling Strategies, IWOMP 2008, Springer LNCS.

\bibitem{elnozahy08} E.N.M. Elnozahy, R. Bianchini et al., System Resilience at Extreme Scale, DARPA technical report, 2008.

\bibitem{ferreira11} K. Ferreira, J. Stearly, J.H. Laros III, R. Oldfield, K. Pedretti, R. Brightwell, R. Riesen, P.G. Bridegs, D. Arnold, 
Evaluating the Viability of Process Replication Reliability for Exascale Systems, SC'11, ACM Press.

\bibitem{frigo98} M. Frigo, C.E. Leiserson, K.H. Randall, The Implementation of the Cilk-5 Multithreaded Language, PLDI'98, ACM Press.

\bibitem{geist11} A. Geist, The Path to Exascale: What is the monster in the closet?, Architectures I Workshop, 2011.

\bibitem{guo09} Y. Guo, R. Barik, R. Raman, V. Sarkar, Work-First and Help-First Scheduling Policies for Terminally Strict Parallel Programs, IPDPS'09, IEEE.

\bibitem{hennessy11} J. Hennessy, D. Patterson, Computer architecture: A quantitative approach, 5th ed., Morgan Kaufman publishers, 2011.

\bibitem{intel11} Intel 64 and IA-32 Architectures Software Developer's Manual, Vol.\ 3: System Programming Guide, Intel, Dec. 2011.

\bibitem{kleen09} A. Kleen, Ongoing evolution of Linux x86 machine check handling, LinuxCon 2009.

\bibitem{kleen10} A. Kleen, mcelog: Memory Error Handling in User Space, Linux-Kongress, 2010.

\bibitem{kruijf12} M. de Kruijf, K. Sankaralingam, S. Jha, Static Analysis and Compiler Design for Idempotent Processing, PLDI'12, ACM Press.

\bibitem{kwok99} Y.-K. Kwok, I. Amhad, Static scheduling algorithms for allocating directed task graphs to multiprocessors, ACM Computing Surveys, 31(4), 1999.

\bibitem{lea00} D. Lea, A Java Fork/Join Framework, Java'00, ACM Press.

\bibitem{michael09} M.M. Michael, M.T. Vechev, V.A. Saraswat, Idempotent Work Stealing, PPoPP'09, ACM Press.

\newpage 

\bibitem{mukherjee08} S. Mukherjee, Architecture Design for Soft Errors, Morgan Kaufman Publishers, 2008.

\bibitem{reinders07} J. Reinders, Intel Threading Building Blocks, O'Reilly, 2007.

\bibitem{rivers09} J.A. Rivers and P. Kudva, Reliability Challenges and System Performance at the Architecture Level, IEEE Design \& Test of Computers, November/December 2009.

\bibitem{schroeder07} B. Schroeder, G. Gibson, Understanding failures in petascale computers, Journal of Physics: Conf. Ser. 78, 2007.

\bibitem{shye09} A. Shye, J. Blomstedt, T. Moseley, V.J. Reddi, D.A. Connors, PLR: A Software Approach to Transient Fault Tolerance for Multicore Architectures, IEEE Transactions on Dependable and Secure Computing, 6(2), April-June 2009.

\bibitem{ullman75} J.D. Ullman, NP-complete scheduling problem, Journal of Computer and System Sciences, 10, 1975.

\end{thebibliography}
\end{document}